\documentclass[letter,11pt]{article}%
\usepackage{xfrac}
\usepackage{amsmath}
\usepackage{amsfonts}
\usepackage{amssymb}
\usepackage{graphicx}
\usepackage{color}
\usepackage{rotating}
\usepackage{natbib}
\usepackage{enumitem}
\usepackage[height=9in,left=1in,right=0.75in,bottom=1in]{geometry}
\usepackage[breaklinks=true,bookmarksopen=true,colorlinks=true,citecolor=blue]%
{hyperref}

\usepackage{threeparttable}

\usepackage{graphicx}
\usepackage{subcaption}
\usepackage{indentfirst} 

\providecommand{\U}[1]{\protect\rule{.1in}{.1in}}
\newtheorem{theorem}{Theorem}[section]
\usepackage{bbm}

\newtheorem{assumption}{Assumption}

\newtheorem{remark}{Remark}[section]

\makeatletter
\renewcommand{\thetheorem}{\thesection.\arabic{theorem}}
\@addtoreset{lemma}{Appendix B} \makeatother

\hypersetup{pdftitle={Alegre and Escanciano}, pdfsubject={Robust Minimum Distance Inference}, pdfauthor={Joan Alegre and Juan Carlos Escanciano}, pdfkeywords={Minimum Distance Estimation; 
Partial identification; Robust inference} }

\begin{document}

\title{Robust Minimum Distance Inference in Structural Models \thanks{Research funded by Ministerio de Ciencia e Innovación, grant ECO2017-86675-P, MCI/AEI/FEDER/UE, grant PGC 2018-096732-B-100, grant PID2021-127794NB-I00 and Comunidad de Madrid, grants EPUC3M11 (VPRICIT) and H2019/HUM-589.} }

\author{
   Joan Alegre\\
  \text{Universidad Carlos III de Madrid}
  \and
  Juan Carlos  Escanciano \\
  \text{Universidad Carlos III de Madrid} 
} 

\date{\today}

\maketitle

\begin{abstract}
This paper proposes minimum distance inference for a structural parameter of
interest, which is robust to the lack of identification of other structural
nuisance parameters. Some choices of the weighting matrix lead to
asymptotic chi-squared distributions with degrees of freedom that can be
consistently estimated from the data, even under partial identification. In
any case, knowledge of the level of under-identification is not required. We study the power
of our robust test. Several examples show the wide applicability of the procedure and a Monte
Carlo investigates its finite sample performance. Our identification-robust inference method can be applied to make inferences on both calibrated (fixed) parameters and any other structural parameter of interest. We illustrate the method's usefulness by applying it to a structural model on the non-neutrality of monetary policy, as in \cite{nakamura2018high}, where we empirically evaluate the validity of the calibrated parameters and we carry out robust inference on the slope of the Phillips curve and the information effect.

\vspace{2mm}

\begin{description}
\item[Keywords:] Minimum Distance Estimation; Partial Identification; Robust inference.

\item[\emph{JEL classification:} C10, C12, C16] 

\end{description}
\end{abstract}

\vfill\pagebreak

\section{Introduction}
Minimum distance inference is commonly used in structural econometric models,
with applications varying from dynamic and static games of imperfect
information (see, e.g., \cite{pesendorfer2008asymptotic}), dynamic
stochastic general equilibrium (DSGE) models (see, e.g., \cite{rotemberg1997optimization}, \cite{amato2003estimation}, and \cite{christiano2005nominal}), to any economic model estimated by simulated-based methods such as Simulated Method of Moments (SMM) or Indirect Inference (II) (see, e.g., \cite{nakamura2018high}). Two key assumptions in the literature are
that structural parameters are point-identified and the corresponding Jacobian matrix is of full column rank. In this paper, we relax these two
conditions and propose inference for a structural parameter of interest
$\beta$ that is robust to lack of under-identification of other nuisance structural parameter $\alpha$. Moreover, it is not necessary to have prior knowledge of the degree of identification of the vector $\alpha$. There is
substantial theoretical and empirical evidence that structural parameters in
econometric models are generally not identified, see the references below. However, the level of underidentification is often unknown to the researcher. We propose an inference method that is robust to such knowledge and also to singularities in the optimal weighting matrix. Therefore, the proposed method provides a simple robust alternative to the standard
econometric practice. Notably, our approach does not require the identification of the $\beta$ parameter either, as the restricted model is used to compute the test statistic following a Lagrange multiplier or score-type approach. As a result, this testing procedure is particularly well-suited to make inferences on calibrated (fixed) parameters in structural macroeconomic and microeconomic applications. We refer to this application of our method as testing the validity of calibrated parameters.

In structural models estimated by minimum distance methods the degree of identification of the structural parameters is often unknown. This is so because of a lack of an analytic solution for the Jacobian (see, e.g., \cite{mcfadden1989method} on SMM or \cite{gourieroux1993indirect} on II), or because the rank of the Jacobian depends on an unknown population parameter (see, e.g, \cite{pesendorfer2008asymptotic} on Dynamic Games,  or \cite{kalouptsidi2021identification} on counterfactuals in Dynamic Discrete Choice models). In this paper, we present an inference method for a parameter of interest that is robust to the lack of identification of the nuisance parameter, both under the null and under the alternative. Importantly, our method does not require any prior knowledge about the unknown degree of identification of the model, which is unknown but identified from data. Additionally, our proposed method has the secondary benefit of providing an estimation of the degree of identification of the model.

We describe our setting. Let $\theta\in\Theta\subset\mathbb{R}^{m}$ be a reduced-form parameter which is identified from the data distribution $F_{0}$ as $\theta_{0}$.
Let $\alpha\in\mathcal{A}\subset\mathbb{R}^{q}$ and $\beta\in\mathcal{B}%
\subset\mathbb{R}^{p}$ be structural parameters. Structural
parameters are related to reduced-form parameters through the mapping $g : \Theta \times \mathcal{A} \times \mathcal{B} \xrightarrow{} \Theta$ such that 
\begin{equation}
    \theta_{0}=g(\theta_{0},\alpha,\beta).
\end{equation}
This paper proposes a test for the
hypothesis
\begin{equation}\label{null}
    H_{0}:\beta=\beta_{0}\qquad vs\qquad H_{1}:\beta\neq\beta_{0},
\end{equation}
where $\beta_{0}$ is a known fixed value, and we allow for the nuisance structural parameter $\alpha$ to be partially
identified under the null, i.e.,  the set
\[
\mathcal{A}_{0}:=\left\{  \alpha\in\mathcal{A}\subset\mathbb{R}^{q}:\theta
_{0}=g(\theta_{0},\alpha,\beta_{0})\right\}
\]
is not necessarily a singleton. We assume that when the null hypothesis is true there exists an $\alpha_{0} \in \mathcal{A}_{0}$, i.e, $\mathcal{A}_{0} $ is non-empty. We also assume there exists an asymptotically normal
estimator for $\theta_{0},$ say $\widehat{\theta},$ satisfying
\[
\sqrt{n}\left(  \widehat{\theta}-\theta_{0}\right)  \rightarrow_{d}N\left(
0,\Sigma\right)  ,
\]
where $\Sigma$ might be a singular matrix. Consider a standard minimum distance (MD) test statistic%
\begin{equation} \label{statistic_F}
    \widehat{F}(\beta_{0}):=\min_{\alpha\in\mathcal{A}}n\left(  \widehat{\theta
}-g(\widehat{\theta},\alpha,\beta_{0})\right) ^{\prime}\widehat{W}\left(
\widehat{\theta}-g(\widehat{\theta},\alpha,\beta_{0})\right)
\end{equation}
where $\widehat{W}$ is a consistent estimator for a positive semi-definite symmetry
non-stochastic matrix $W.$ Assume $g(\theta,\alpha,\beta_{0})$ is twice
continuously differentiable at $\theta_{0}\in\Theta\ $and $\alpha_{0}%
\in\mathcal{A}_{0},$ and define the matrices $\nabla_{\theta}g_{0}:=\partial
g(\theta_{0},\alpha_{0},\beta_{0})/\partial\theta^{\prime}$ and $\nabla
_{\alpha}g_{0}:=\partial g(\theta_{0},\alpha_{0},\beta_{0})/\partial
\alpha^{\prime}.$ Let $r_{\alpha}$ and $r_{\Sigma}$ denote the ranks of $\nabla_{\alpha}g_{0}$ and $\Sigma$, respectively. We assume the rank of the variance and covariance matrix is larger than the rank of the Jacobian, i.e., $r_{\Sigma}>r_{\alpha}$. Under the
regularity conditions below, $r_{\alpha}$ does not depend on the point $\alpha
_{0}\in\mathcal{A}_{0}$ where it is evaluated. Let $I_{m}$ be the $m$-dimensional identity matrix. The main result of the paper shows that
under regularity conditions and $H_{0},$ if $W=\left[  (I_{m}-\nabla_{\theta
}g_{0})\Sigma(I_{m}-\nabla_{\theta}g_{0})^{\prime}\right]  ^{\dagger}$, where $A^{\dagger}$ denotes the Moore-Penrose generalized inverse of the
matrix $A$, then
\begin{equation}
\widehat{F}(\beta_{0})\rightarrow_{d}\chi_{r_{\Sigma}-r_{\alpha}}^{2},\label{null}%
\end{equation}
 where $\chi_{d}^{2}$
denotes a chi-squared distribution with $d$ degrees of freedom. This result
includes the case where $\alpha_{0}$ is point identified and $r_{\alpha}=q$ (see \cite{gourieroux1995testing}
). The more general case discussed here allows for
lack of identification of $\alpha_{0}$ and $r_{\alpha}<q.$ The convergence in
(\ref{null}) suggests a feasible $\tau\%$ nominal test with critical region
$\widehat{F}(\beta_{0})>\chi_{\widehat{r}_{\Sigma}-\widehat{r}_{\alpha},1-\tau}^{2},$ where $\widehat{r}_{\Sigma}-\widehat{r}_{\alpha}$
is a consistent estimator of $r_{\Sigma}-r_{\alpha}$ and $\chi_{d,\tau}^{2}$ denotes the $\tau
-$quantile of $\chi_{d}^{2}.$
The proposed method applies general results on the second-order
differentiability properties of the mapping%
\[
v(\theta):=\min_{\alpha\in\mathcal{A}}\left(  \theta-g(\theta,\alpha,\beta
_{0})\right)  ^{\prime}W\left(  \theta-g(\theta,\alpha,\beta_{0})\right)  ,
\]
at $\theta=\theta_{0},$ see \cite{shapiro1986asymptotic}. When $\alpha_{0}$ is a regular
point (see definition below) and other regularity conditions hold, $v(\theta)$
is twice continuously differentiable and it can be locally approximated by the
projection norm onto a tangent linear space of dimension $r_{\alpha}.$ That is, the
$q-$dimensional manifold
\[
\Theta_{0}:=\left\{  \theta\in\Theta\subset\mathbb{R}^{m}:\theta=g(\theta
,\alpha,\beta_{0}), \hspace{0.25cm} \textit{for some} \hspace{0.25cm}\alpha\in\mathcal{A}\subset\mathbb{R}^{q}\right\}
\]
has a tangent space of dimension $r_{\alpha}$ at $\theta_{0},$ which is generated by
the columns of $\nabla_{\alpha}g_{0}.$ This result does not require
identification of $\alpha_{0}$ or full column rank of $\nabla_{\alpha}g_{0}.$
The asymptotic distribution of $\widehat{F}(\beta_{0})$ then follows from a
standard application of the Delta-Method, after noticing that $\widehat
{F}(\beta_{0})=v(\widehat{\theta})+o_{P}(1)$.

One important novelty of our method is that it does not require any prior knowledge about the degree of identification of the model, i.e, $r_{\alpha}$, as long as $r_{\Sigma}>r_{\alpha}$. Such lack of knowledge about the degree of identification arises whenever the Jacobian matrix depends on the population parameter or there is no analytic solution for it. In SMM, for instance, no analytical solutions of the Jacobian are available, though it can be approximated. In structural models such as Dynamic Games or Dynamic Discrete Choice models, a closed-form solution for $r_{\alpha}$ is available, but it does depend on the population value of $\theta_{0}$, which is unknown. We estimate the rank of $\Sigma$ and $\nabla_{\alpha}g_{0}$ using a hard thresholding approach.

A prototypical application of our results is for testing the validity of calibrated parameters in structural models. This refers to testing for $H_{0}:\beta=\beta_{0}$ for calibrated values $\beta_{0}$ of a structural parameter. ``Validity'' corresponds to a lack of rejection by our robust test. Another application of our results could be for the construction of robust confidence intervals for a structural parameter of interest by inverting our robust test. We evaluate the finite sample performance of our tests (size and power) in the context of two sets of Monte Carlo experiments, one for a static Bayesian entry game and another for the New-Keynesian model in \cite{nakamura2018high}. The Monte Carlo results confirm the robustness of our method, relative to traditional non-robust tests based on asymptotic normality under point-identification (t-tests), and show that estimation of the level of under-identification does not have an impact on the performance of the robust test.

We apply our robust method to study both the validity of the calibrated parameters and the inference on the slope of the Phillips curve and the information effect in a simplified version of \cite{nakamura2018high}. We find their calibration to be valid according to our criteria, and obtain robust confidence intervals for the slope of the Phillips curve and the information effect, which are not only shorter than those obtained by the classical t-test but also shorter than those suggested by the nonparametric bootstrap. The identification-robust inference seems to be well motivated in this application given that the Jacobian for nuisance structural parameters is of reduced rank and the near singularity of the asymptotic variance matrix of reduced form parameters.

The paper is organized as follows: after this introduction and a short
literature review, Section \ref{Examp} discusses several examples of
structural econometric models where our results apply. Section \ref{MD}
provides the asymptotic null distribution and power theory for the proposed test. Section \ref{implem} discusses the implementation.
Section \ref{MonteCarlo} investigates the finite sample performance of the proposed
methods, while Section \ref{AP} illustrates our method in the application of \cite{nakamura2018high}. Finally, Section \ref{concl} concludes. The main proofs of Section
\ref{MD} are gathered into \ref{Main_Proofs}.

\section{Literature Review\label{LR}}This paper proposes an identification-robust inference method for parameters of interest allowing for nuisance parameters to be unidentified both under the null and under the alternative hypothesis, without prior knowledge of the model's identification degree.

Our problem can be placed in the literature of identification-robust inference of a parameter of interest when the nuisance parameters are not identified under the null.  Some examples of this literature are \cite{davies1977hypothesis}, \cite{engle1984wald}, \cite{andrews1994optimal}, \cite{hansen1996inference}, \cite{stinchcombe1998consistent}, \cite{conniffe2001score}. Such literature indexes the test statistic by the nuisance parameter, and then, finds the supremum of all possible test statistics. The distribution of the supremum test statistic is not standard, and they use Monte Carlo or resampling methods to find critical values for the test. Our method leads to critical values that are known up to the degrees of freedom, which can be estimated from data without resampling.
Our environment allows the model to be partially identified. Inference on partially identified models has been treated by \cite{horowitz1998censoring}, \cite{horowitz2000nonparametric}, \cite{chernozhukov2002inference}, \cite{imbens2004confidence}, \cite{romano2008inference}, \cite{beresteanu2008asymptotic}, \cite{chernozhukov2007estimation}, \cite{bugni2010bootstrap}, \cite{bugni2016comparison}, and many others. As a common trait, such literature has focused both on constructing confidence intervals that contain the whole identified set and developing confidence intervals for the true value. Most of the methods rely on subsampling or bootstrap to carry out inference.

This paper also relates to the literature on testing with a singular information matrix, such as \cite{satorra1989alternative}, \cite{satorra1992asymptotic}, \cite{andrews1987asymptotic},  \cite{ravikumar2000robust} or \cite{dufour2016rank}, among many others. Our paper mainly differs from this literature in the assumption of point identification of the nuisance parameter.

Inference under weak identification is also related to our problem. Some papers in this literature are  \cite{stock2000gmm,andrews2012estimation, andrews2013maximum,andrews2015maximum, andrews2016conditional, andrews2016geometric, han2019estimation, andrews2019identification}; or \cite{lee2022robust}.
\cite{kleibergen2005testing} assumes the full column rank of the Jacobian for the nuisance parameter. \cite{andrews2016geometric} gives finite sample bounds on the distribution of minimum statistics used in minimum distance models. They assume the full column rank of the Jacobian. Moreover, in their framework the model $g(\cdot)$ does not depend on the reduced-form parameter $\theta_{0}$. \cite{andrews2019identification} develop a singular weighting matrix robust test. They also assume identification of the nuisance parameter. 
\cite{andrews2012estimation} develops an estimation and inference method that is robust to lack of identification (and thus weak identification) of nuisance parameters. There are two main differences with our paper. First, to use their method, knowledge about which parameters can be identified is required, while we are agnostic about the identification problem. Second, the objective function does not depend on the nuisance parameter when it is not identified, while we allow the objective function to depend on the nuisance parameter even if it is not identified. \cite{han2019estimation} develops estimation and inference in a GMM framework, with a Jacobian that is not full column rank. \cite{han2019estimation} works on the \cite{andrews2012estimation} environment, and therefore, prior knowledge of the identification problem is required. Moreover, the objective function cannot depend on unidentified nuisance parameters.

\cite{antoine2022identification} develop an identification-robust inference of a structural parameter of interest using a minimum distance estimation method. The main difference with our paper is that we allow for under-identification of the nuisance parameter. \cite{antoine2022identification} also relies on bootstrap methods to find the critical values of the test. We avoid using bootstrap, making our procedure computationally less expensive for a large variety of complex structural models.
\section{Examples\label{Examp}}

\subsection{Estimation of Games}

The proposed method is motivated by the application of dynamic games by
\cite{pesendorfer2008asymptotic}. These authors suggested the MD
estimator%
\begin{equation}
\widehat{\alpha}:=\arg\min_{\alpha\in\mathcal{A}}n\left(  \widehat{\theta
}-g(\widehat{\theta},\alpha)\right)  ^{\prime}\widehat{W}\left(
\widehat{\theta}-g(\widehat{\theta},\alpha)\right)  ,\label{MDG}%
\end{equation}
where $\widehat{\theta}$ is a consistent estimator for the ex-ante choice
probabilities $\theta_{0}$ and  $g$ is the best response mapping, linking the
structural parameter $\alpha_{0}$ to the choice probability $\theta_{0}$. In
equilibrium $\theta_{0}=g(\theta_{0},\alpha_{0}).$ Much of the literature
assumes $\alpha_{0}$ is uniquely identified from the equilibrium conditions.
However, there is theoretical evidence that these models are not identified.
Non-identification results have been shown in the seminal paper by \cite{rust1994structural} and \cite{magnac2002identifying} for single-agent models and by \cite{pesendorfer2008asymptotic} for multiple-agent models. Here, we present a methodology that can be
applied to partially identified models, and therefore, has wider applicability
than existing procedures. 
\cite{pesendorfer2008asymptotic} show how many popular estimators fall
under the class (\ref{MDG}). Specifically, they show that the moment estimator
of \cite{hotz1993conditional} or the pseudo-maximum likelihood estimator of
\cite{aguirregabiria2002identification} are asymptotically equivalent to estimators in
the class (\ref{MDG}) for specific choices of the plim of $\hat{W}$, $W.$ Here, we allow these choices
as special cases, while permitting the structural parameters to be
under-identified. 

\subsection{Estimation of Dynamic Stochastic General Equilibrium Models}

In this application $\theta_{0}$ is a vector of impulse response functions
obtained from a Structural Vector Autoregression model, and $g(\cdot)$ is the link
between structural parameters and reduced-form parameters. MD estimation in
this context has been proposed by \cite{rotemberg1997optimization}, \cite{amato2003estimation}, and \cite{christiano2005nominal} , among many others. In this application, $g$ does not depend on $\theta_{0}$.

\subsection{Estimation using Simulated Based Methods}
Simulated-based methods such as SMM and II also fit our environment, after an appropriate modification that we give below in Section \ref{smm}. Moreover, due to the nature of their problem, the degree of identification of the model is unknown. Furthermore, simulated-based methods are computationally demanding, making bootstrap and subsampling inference not feasible computationally. Therefore, our method addresses two main problems in this literature. It is agnostic about the degree of identification, while it is also computationally feasible.

\subsubsection{Estimation with Simulated Method of Moments}\label{smm}
This method is introduced in the seminal papers by  \cite{mcfadden1989method}, \cite{pakes1989simulation},  \cite{duffie1990simulated}, and \cite{lee1991simulation}.

	Assume we observe a sample of random variables $X_{i}$ of size $n$, where $X_{i}$ follows a distribution $ F(\cdot)$. Moreover, suppose we have a structural economic model, indexed by parameters $(\alpha,\beta)$ that generates a simulated random variable $\tilde{X}_{j}(\alpha,\beta) \sim F_{\alpha,\beta}$. For a given measurable function: 
	\begin{equation}
		h: \mathcal{X} \xrightarrow{} \Theta,
	\end{equation}
	where $\mathcal{X}$ is the domain of  $X_{i}$, let
	\begin{equation} \label{reduced}
		\theta_{0} = \int h(x_{i})d F(x_{i}).
	\end{equation}
	The reduced-form parameter $\theta_{0}$ is estimated using the sample analog in (\ref{reduced}).
 In SMM, the moment $g(\cdot)$ is defined by
	\begin{equation} \label{numerical}
		g(\alpha,\beta) := \int h(\tilde{x}_{i})d F_{\alpha,\beta}(\tilde{x}_{i}),
	\end{equation}
	which is not analytically known. We use Monte Carlo, and compute
	\begin{equation} \label{simulated}
		\hat{g}(\alpha, \beta) \equiv \frac{1}{B}\sum_{j=1}^{B} h(\tilde{x}_{j}(\alpha,\beta)),
	\end{equation}
	where $\{x_{j}(\alpha,\beta)\}$ are $B$ random draws from $F_{\alpha,\beta}$. The SMM minimizes a quadratic objective function that depends on the distance between the estimated reduced-form parameter (\ref{reduced}) and the simulated moment condition (\ref{simulated}), i.e.,
	
	\begin{equation} \label{smm_objective}
		\tilde{F}(\beta_{0}):=\min_{\alpha\in\mathcal{A}}n\left(  \widehat{\theta
		}-\hat{g}(\alpha, \beta_{0})\right)  ^{\prime}\widehat{W}\left(
		\widehat{\theta}-\hat{g}(\alpha, \beta_{0})\right), 
	\end{equation}
where $\widehat{W}$ will be the estimated optimal weighting matrix of this minimum distance problem. By choosing $B$ significantly large, one shows that the numerical approximation of (\ref{simulated}) has no impact on the asymptotic distribution of (\ref{smm_objective}), 
	\begin{equation}
		\hat{F}(\beta_{0}) = \tilde{F}(\beta_{0})+o_{P}(1),
	\end{equation}
	and hence, SMM fits our framework with a $g(\cdot)$ given by (\ref{numerical}) that does not depend on $\theta$.

\section{Minimum Distance Inference\label{MD}}
This section investigates the asymptotic null distribution of $\widehat
{F}(\beta_{0}).$ The first condition requires the asymptotic normality of the
reduced-form estimator. The test statistic was defined in equation (\ref{statistic_F}) as
\begin{equation*}
    \widehat{F}(\beta_{0}):=\min_{\alpha\in\mathcal{A}}n\left(  \widehat{\theta
}-g(\widehat{\theta},\alpha,\beta_{0})\right) ^{\prime}\widehat{W}\left(
\widehat{\theta}-g(\widehat{\theta},\alpha,\beta_{0})\right).
\end{equation*}

\begin{assumption}%
\label{red}The parameter $\theta_{0}$ belongs to the interior of
$\Theta\subset\mathbb{R}^{m}.$ Under $H_{0},$ $\sqrt{n}\left(  \widehat
{\theta}-\theta_{0}\right)  \rightarrow_{d}N\left(  0,\Sigma\right)  ,$ where
$\Sigma$ might be singular.%

\end{assumption}%

%

\begin{assumption}%

\label{differentiability}The function $g(\theta,\alpha,\beta_{0})$ is twice
continuously differentiable at $\theta_{0}\in\Theta\ $and $\alpha_{0}%
\in\mathcal{A}$.%

\end{assumption}%

\begin{assumption}  \label{correct specification} The model is correctly specified,

\begin{equation}
    \theta_{0}=g(\theta_{0},\alpha,\beta) \hspace{0.3cm} \textit{for some} \hspace{0.1cm} \beta \in \mathcal{B} \hspace{0.1cm} \textit{and} \hspace{0.1cm }\alpha \in \mathcal{A},
\end{equation}
\end{assumption}

These assumptions are standard in the literature on MD inference. We say the
point $\alpha_{0}\in\mathcal{A}$ is locally regular if belongs to the interior
of $\mathcal{A}$ and the Jacobian matrix $\nabla_{\alpha}g(\alpha):=\partial
g(\theta_{0},\alpha,\beta_{0})/\partial\alpha^{\prime}$ has the same rank as
$\nabla_{\alpha}g_{0}\equiv\nabla_{\alpha}g(\alpha_{0})$, say $r_{\alpha},$ for every
$\alpha$ in a neighborhood of $\alpha_{0}.$ We say the point $\alpha_{0}%
\in\mathcal{A}$ is regular if it is locally regular and there exist neighborhoods
$\mathcal{U}$ and $\mathcal{V}$ of $\alpha_{0}$ and $\theta_{0},$
respectively, such that $\Theta_{0}\cap\mathcal{V}=g(\theta_{0},\mathcal{U}%
,\beta_{0}),$ where $\Theta_{0} :=\left\{  \theta\in\Theta\subset\mathbb{R}^{m}:\theta=g(\theta
,\alpha,\beta_{0}), \hspace{0.25cm} \textit{for some} \hspace{0.25cm}\alpha\in\mathcal{A}\subset\mathbb{R}^{q}\right\} $.

%
\begin{assumption}%
\label{reg} (i) $\mathcal{A}$ is a compact subspace of $\mathbb{R}^{q}$; (ii)
the identified set $\mathcal{A}_{0}$ is connected$;$ and (iii) all elements of
$\mathcal{A}_{0}$ are locally regular.%
\end{assumption}%
%
\begin{assumption} \label{wei}%
\label{W}$\widehat{W}=W+o_{P}(1)$, with $W$ a positive semi-definite and symmetric
non-stochastic matrix.%
\end{assumption}%
\begin{assumption} \label{rank overparam} $W$ is positive definite over the space spanned by the columns of $\nabla_{\alpha}g_{0}$, i.e., 
\begin{equation}
Rank(\nabla_{\alpha} g_{0}'W\nabla_{\alpha}g_{0}) =  
Rank(\nabla_{\alpha}g_{0}).
\end{equation}
\end{assumption}

\begin{assumption} 
\label{Non-deficiency} The matrix $I_{m}-\nabla_{\theta}g_{0}$ is non-singular.
\end{assumption}

\bigskip
Assumption \ref{reg}$(i)$  is standard and it can be relaxed at the cost of
introducing further assumptions on $g.$ A subset $S$ of a topological space $\mathcal{X}$ is said to be connected whenever $S$ cannot be expressed as the disjoint union of two non-empty open subsets of $\mathcal{X}$.  When Conditions \ref{reg}$(i)$, \ref{reg}$(ii)$  and \ref{reg}$(iii)$ do not hold, our test controls the size, although it might become conservative, see Remark \ref{conservative test}. Assumptions \ref{wei} and \ref{rank overparam} are standard in the literature. Assumption \ref{Non-deficiency} is a local identification condition for reduced form parameters.


\begin{theorem} \label{distr} Let Assumptions \ref{red} -- \ref{Non-deficiency} hold, and set
$W=\left((I_{m}-\nabla_{\theta}g_{0})\Sigma
(I_{m}-\nabla_{\theta}g_{0})^{\prime}\right)^{\dagger}$. Then,

\begin{equation}
    \widehat{F}(\beta_{0}) \xrightarrow{d}  \chi_{r_{\Sigma}-r_{\alpha}}^{2}
\end{equation}

\end{theorem}



\begin{remark}
If the gradient $\nabla_{\alpha}g$ is not available in closed form, one can
use finite-difference approximations, as justified in e.g. \cite{hong2015extremum}. For simplicity, we assume that a closed form of the derivatives is available.
\end{remark}
\begin{remark}\label{conservative test}
    If Assumptions \ref{red} -- \ref{correct specification}, Assumption \ref{W} -- \ref{Non-deficiency} hold, and $W=\left(I_{m}-\nabla_{\theta
}g_{0})\Sigma(I_{m}-\nabla_{\theta}g_{0})^{\prime}\right)^{\dagger}$,  then 
    \begin{equation}
        \limsup_{n\xrightarrow{}\infty}\Pr\{\widehat{F}(\beta_{0})\geq c\} \leq Pr\{\chi^{2}_{r_{\Sigma}-r_{\alpha}}\geq c\}.
    \end{equation}
See  Theorem $3.3$ in \cite{shapiro1986asymptotic}. This result gives a conservative test for cases where Assumption \ref{reg} does not hold. Violations of Assumption \ref{reg}$(ii)$ might occur, for example, when $\mathcal{A}_{0}$ is a discrete set. Violations of Assumption \ref{reg}$(iii)$ might occur whenever the model is reparametrized to avoid boundary solutions. For example, let $g(\alpha) = \alpha$ and $\mathcal{A}=\{\alpha \in \mathbb{R} \hspace{0.1cm}| \hspace{0.1cm}\alpha \geq 0\}$, we might reparametrize the model using the mapping $\alpha \xrightarrow{} \alpha^{2}$. If $\alpha_{0}=0$, $\alpha_{0}$ is not a regular point of $\mathcal{A}_{0}$.
\end{remark}
If $W=\left((I_{m}-\nabla_{\theta}g_{0})\Sigma
(I_{m}-\nabla_{\theta}g_{0})^{\prime}\right)^{\dagger}$, then the limiting
distribution depends on the data-generating process only through the rank of
the matrix $\nabla_{\alpha}g_{0}.$ In this case, it is natural to consider the
sample analog%
\[
\widehat{r}_{\alpha}=rank\left[  \nabla_{\alpha}g(\widehat{\theta},\widehat{\alpha
},\beta_{0})\right]  ,
\]
where $\widehat{\alpha}$ solves the optimization problem
\begin{equation}
    \label{ridge}
\widehat{\alpha}=\arg\min_{\alpha\in\mathcal{A}}n\left(  \widehat{\theta
}-g(\widehat{\theta},\alpha,\beta_{0})\right)  ^{\prime}\widehat{W}\left(
\widehat{\theta}-g(\widehat{\theta},\alpha,\beta_{0})\right)  +\lambda
_{n}\left\vert \alpha\right\vert ^{2},
\end{equation} 
and $\lambda_{n}$ is a penalization parameter such that $\lambda_{n}\downarrow0$.

Furthermore, to satisfy Assumption \ref{wei} whenever  $W=\left((I_{m}-\nabla_{\theta}g_{0})\Sigma
(I_{m}-\nabla_{\theta}g_{0})^{\prime}\right)^{\dagger}$, an estimator based on the analog principle of $W$ might be inconsistent. This is because the Moore-Penrose g-inverse is not continuous. \cite{stewart1969continuity} gives necessary and sufficient conditions to estimate consistently a Moore-Penrose g-inverse. A detailed discussion on how to construct a consistent estimator of $W$ is provided in Section \ref{consistent_MOORE}. 
%
\begin{assumption}
    
\label{West}(i) $\widehat{\Sigma}=\Sigma+o_{P}(1);$ (ii)
$\alpha_{0}=\arg\min\left\{  \left\vert \alpha\right\vert :\alpha
\in\mathcal{A}_{0}\right\}  $ is unique $;$ and (iii) $\lambda_{n}\rightarrow0$ as
$n\rightarrow\infty.$%
\end{assumption}%

Henceforth, to save notation we denote $d \equiv  r_{\Sigma}-r_{\alpha}$ and $\widehat{d} \equiv  \widehat{r}_{\Sigma}-\widehat{r}_{\alpha}$.
Consider the oracle test $\tilde{\phi}_{\tau}=1(\widehat{F}(\beta_{0}%
)>\chi_{d,1-\tau}^{2}),$  the one that uses the true degree's of freedom $d$, where $1(A)$ denotes the indicator
function of the event $A$ (1 if $A$ holds, and 0 otherwise). Define the
asymptotic power function $\pi_{\tau}(\beta)\equiv \lim_{n\rightarrow\infty
}\mathbb{P}_{\beta}\left(  \tilde{\phi}_{\tau}=1\right)  $, where $\mathbb{P}%
_{\beta}$ denotes the underlying probability when the true parameter is
$\beta.$  Call feasible test $\hat{\phi}_{\tau} = 1(\widehat{F}(\beta_{0}%
)>\chi_{\widehat{d}}^{2})$, i.e, the test using the estimated degree of freedom $\widehat{d}$.

\begin{theorem}\label{feasible equal oracle} If $lim_{n\rightarrow\infty
}\mathbb{P}_{\beta}(\widehat{d} = d)= 1$ then, for all $\beta$,
\begin{equation}
    \lim_{n\rightarrow\infty
}\mathbb{P}_{\beta}\left(  \hat{\phi}_{\tau}=1\right)  = \pi_{\tau}(\beta).
\end{equation}
\end{theorem}
The next theorem justifies the proposed feasible test. Define the set
\[
\mathcal{B}_{1}:=\left\{  \beta_{1}\in\mathcal{B}\subset\mathbb{R}^{p}%
:\min_{\alpha\in\mathcal{A}}\left( g(\theta_{0},\alpha_{1},\beta_{1})-g(\theta
_{0},\alpha,\beta_{0})\right)^{\prime}W\left( g(\theta_{0},\alpha_{1},\beta_{1})-g(\theta
_{0},\alpha,\beta_{0})\right) >0\right\}.
\]
where $\alpha_{1} \equiv \alpha_{1}(\beta_{1})$ is the minimum norm solution of the problem
\begin{equation}
     \min_{\alpha\in\mathcal{A}}\left(  \theta_{0}-g(\theta
_{0},\alpha,\beta_{1})\right)^{\prime}W\left( \theta_{0}-g(\theta
_{0},\alpha,\beta_{1})\right).
\end{equation}

\begin{theorem} \label{power} \label{feasible}Let Assumptions \ref{red} -- \ref{West}  hold.
Then, for all $\tau\in(0,1),$ $\pi_{\tau}(\beta_{0})=\tau,$ whereas $\pi
_{\tau}(\beta)=1$ for all $\beta\in\mathcal{B}_{1}$.
\end{theorem}
\begin{remark}
Theorem \ref{feasible} shows that the feasible test controls size
asymptotically, and it is consistent for all $\beta
    \in\mathcal{B}_{1}$. If $\beta=\beta_{1} \notin \mathcal{B}_{1}$, $\beta_{1}\neq \beta_{0}$, then there is no information in the model to separate $(\alpha_{1},\beta_{1})$ from $(\alpha_{0},\beta_{0})$, where $\alpha_{1}$ is such that $g(\theta_{1},\alpha_{1},\beta_{1})=g(\theta_{0},\alpha_{0},\beta_{0})$. No test has power against such observational equivalent cases.
\end{remark}

Theorem \ref{power} shows that the test have power if differences between $g(\theta_{1},\alpha_{1},\beta_{1})$ and $g(\theta_{0},\alpha_{0},\beta_{0})$ can be detected when using $||x||_{W} = \left(x'Wx\right)^{\frac{1}
{2}}$ as seminorm, i.e, $||g(\theta_{1},\alpha_{1},\beta_{1})-g(\theta_{0},\alpha_{0},\beta_{0})||_{W} > 0$.

\begin{remark}

The test of Theorem \ref{feasible} is straightforward to implement, and it can
be used to construct confidence regions for $\beta_{0}$ by the classical idea
of inverting the test statistic. 
\end{remark}

The next results give sufficient conditions to have local power under Pittman alternatives. Let $\beta_{1n} = \beta_{0} + \frac{\delta}{\sqrt{n}}$, and with some abuse of notation denote the
asymptotic local power function by $\pi_{\tau}(\delta)\equiv \lim_{n\rightarrow\infty
}\mathbb{P}_{\beta_{1n}}\left(  \tilde{\phi}_{\tau}=1\right) $. We call $\delta$ the "direction" of the alternative.

\begin{assumption} \label{differentiability of beta}
$g(\theta,\alpha,\beta)$ is twice
continuously differentiable at $\theta_{0}\in\Theta\ $, $\alpha_{0}%
\in\mathcal{A}_{0}$, and $\beta_{0} \in \mathcal{B}$.%
\end{assumption}
 Define the Jacobian of the parameter of interest as $\nabla
_{\beta}g_{0}:=\partial g(\theta_{0},\alpha_{0},\beta_{0})/\partial
\beta^{\prime}. $

\begin{theorem} \label{power local} Under Assumptions \ref{red} -- \ref{Non-deficiency}$(i)$ and Assumption \ref{differentiability of beta}, the test will have non-trivial local power whenever the direction $\delta$ is not in the null space of the matrix $W\nabla_{\beta}g_{0}$, i.e, 
\begin{equation}
    \pi_{\tau}(\delta)>\tau \iff \delta \notin Null(W\nabla_{\beta}g_{0}).
\end{equation} 

\end{theorem}

\begin{theorem} \label{maximum direction} Under Assumptions \ref{red} -- \ref{Non-deficiency}$(i)$ and Assumption \ref{differentiability of beta}, the direction of maximum local power, i.e. $\delta^{*} = arg\max_{||\delta||_{2}=1} \pi_{\tau}(\delta)$, is given by
\begin{equation}
    \delta^{*} = arg\max_{||\delta||_{2}=1} \delta' \nabla_{\beta}g_{0}'W\nabla_{\beta}g_{0} \delta,
\end{equation}
that is, $\delta^{*}$ is the eigenvector associated to the maximum eigenvalue of the matrix $ \nabla_{\beta}g_{0}'W\nabla_{\beta}g_{0}.$ 
\end{theorem}

\begin{remark} \label{relative weights}
    Since $||\delta^{*}||_{2}=1$,  elements of $\delta_{i}^{*}$ to the squared can be interpreted as the relative importance of the element $\beta_{i}$ in the direction of maximum local power. We call $\sigma^{*}_{i} \equiv (\delta_{i}^{*})^{2}$ relative weights.
\end{remark}

\begin{remark} \label{local_power_eigen}
The value of $k^{*}$, obtained as the solution to the optimization problem
\begin{equation}
k^{*} = \max_{||\delta||_{{2}=1}} \delta' \nabla{\beta}g_{0}'W\nabla_{\beta}g_{0} \delta,
\end{equation}
serves as the non-centrality parameter for the $\chi^{2}(d,k^{*})$ assymptotic distribution associated with $\widehat{F}(\beta_{0})$ under the local alternative. A larger $k^{*}$ value corresponds to a greater level of local power. This value is the maximum eigenvalue of the matrix $ \nabla_{\beta}g_{0}'W\nabla_{\beta}g_{0} $.
\end{remark}

\begin{remark}
By the Rank-Nulity theorem and the result stated in Theorem \ref{power local}, we can determine the dimension of the subspace in $\mathbb{R}^{p}$ that consists of all the directions our test cannot reject with non-trivial local power. This subspace is denoted as $\mathcal{B}{\tau} = \{\delta \in \mathbb{R}^{p} \hspace{0.1cm} : \hspace{0.1cm} W\nabla{\beta}g_{0}\delta = 0\}$. Consequently, the dimension of $\mathcal{B}{\tau}$ is given by $dim(\mathcal{B}{\tau}) = p-Rank(W\nabla_{\beta}g_{0})$, and it can be estimated accordingly.
\end{remark}

\section{Implementation \label{implem}}
This section will illustrate the test algorithm, with additional details provided in the following subsections. 

\begin{enumerate}
    
    \item Fix $\beta=\beta_{0}$.
   
    \item Compute the reduced-form parameter estimator $\widehat{\theta}$.
    \item Compute the estimated asymptotic variance matrix of the estimated reduced-form parameter $\hat{\Sigma}$.
    \item Solve problem (\ref{ridge}) setting $\lambda_{n}$ by generalized cross-validation and $W = \mathbf{I}_{m}$.
    \item Using $\widehat{\alpha}$ and $\widehat{\theta}$ compute  $I_{m}-\widehat{\nabla}_{\theta}g$.
    \item Estimate the rank $\widehat{r}_\Sigma$. Check Subsection \ref{rank_est} to follow a detailed explanation.
    \item Estimate the optimal weighting matrix $W=\left((I_{m}-\nabla_{\theta}g_{0})\Sigma
    (I_{m}-\nabla_{\theta}g_{0})^{\prime}\right)^{\dagger}$ as in Subsection \ref{consistent_MOORE}.
    \item Solve problem (\ref{ridge}) using $\hat{W}$ and update $\widehat{\alpha}$.
    \item Compute $\widehat{F}(\beta_{0}):=n\left(  \widehat{\theta
    }-g(\widehat{\theta},\widehat{\alpha},\beta_{0})\right)  ^{\prime}\widehat{W}\left(
    \widehat{\theta}-g(\widehat{\theta},\widehat{\alpha},\beta_{0})\right) $.
    \item Estimate the rank $\widehat{r}_\alpha$ using the estimator of expression (\ref{est_rank}). Check Subsection \ref{rank_est} to follow a detailed explanation. Compute $\widehat{d} \equiv  \widehat{r}_{\Sigma}-\widehat{r}_{\alpha}$.
    \item Reject null hypothesis at level $\tau$ when $\hat{\phi}_{\tau}=1(\widehat{F}(\beta_{0}%
)>\chi_{\widehat{d},1-\tau}^{2})$.
   
\end{enumerate}
\subsection{Rank estimation} \label{rank_est} \label{soubroutine}
In this subsection, we motivate the simple hard-thresholding method we use to estimate the rank of a matrix. Take $\widehat{M}$ and $M$ as general positive semi-definite symmetric matrices of dimension $m\times m$,  with $\widehat{M} \xrightarrow{p} M$, and $r_{M} = Rank(M)$. The estimator for the rank of $M$ is:
\begin{equation} \label{est_rank}
    \widehat{r}_{M} = \#\{ \widehat{\lambda}_{i} \geq \frac{1}{n^b}, \hspace{0.2cm} i \in 1,...,m \},
\end{equation}
where $\widehat{\lambda}_{i}$ is the $i$th eigenvalue of $\widehat{M}$ and $0<b<1$ is a tuning parameter. Previous literature such as \cite{lutkepohl1997modified} or \cite{dufour2016rank} has used $b= \frac{1}{3}$, but we instead use $b=0.99$. Our choice is motivated by the better performance of our test in the Monte Carlo exercise. We have tuned other values of $b>\frac{1}{2}$ and obtained similar results. Future research will investigate cross validated methods for the choice of $b$.
\begin{assumption} \label{norm_rank} Asymptotic normality of the estimated matrix $\widehat{M}$
    \begin{equation}
    \sqrt{n}\left(Vec(\widehat{M})-Vec(M) \right) \xrightarrow{d} N_{m^{2}}(0,\Omega),
\end{equation}
\end{assumption}
with $s=Rank(\Omega) \leq m^{2}.$


\begin{theorem} \label{rank_estimation_theorem}
    Under Assumption \ref{norm_rank} $\widehat{r}_{M} \xrightarrow{p} r_{M}$.
\end{theorem}
Instead of using spectral cut-off methods, a common alternative is to use a sequential testing procedure to estimate the rank. Statistical tests, such as those proposed by \cite{robin2000tests} or \cite{kleibergen2005testing} can be employed for this purpose. However, this approach requires a consistent estimator of the asymptotic variance matrix, denoted as $\Omega$, of $\sqrt{n}\left(Vec(\widehat{M})-Vec(M) \right)$. Obtaining such an estimator, $\hat{\Omega}$, might be challenging in our context, particularly when a closed-form solution for the matrix $M$ is unavailable. On the other hand, spectral cut-off methods, such as those used by \cite{lutkepohl1997modified}, \cite{dufour2016rank} or \cite{lee2022robust} among others, do not require estimating $\Omega$. Our approach differs from theirs in that we exploit the fact that the estimate $\widehat{\lambda}_{i}$ converges faster than $\sqrt{n}$ to the population eigenvalue $\lambda_{i}$ whenever $\lambda_{i} = 0$.\\
The following Theorem is taken from \cite{robin2000tests}, see the Appendix for a proof.
\begin{theorem} \label{RS proof}
   \label{RS proof}
Under Assumption \ref{norm_rank}, estimated eigenvalues whose population eigenvalue is 0 converge to the population parameter at a rate of $n^{-1}$. Assume population eigenvalues are in descending order, then,
\begin{equation}
\widehat{\lambda}_{i} = O_{P}\Big(\frac{1}{n}\Big) \hspace{1cm} \forall i\in {r_{M}+1,\dots,m}.
\end{equation}

\end{theorem}

\begin{remark}
Since $Rank(\nabla_{\alpha}g_{0}) = Rank(\nabla_{\alpha}g_{0}\nabla_{\alpha}g_{0}')$, Theorem \ref{rank_estimation_theorem} can be used to estimate $\hat{r}_{\alpha}$, and noticing that Assumption \ref{norm_rank} holds whenever $\hat{\theta}$ is asymptotically normally distributed.
\end{remark}

To sum up, we illustrate the algorithm for the estimation of the rank of both $\Sigma$ and $\nabla_{\alpha}g_{0}$.
\begin{enumerate}
    \item Estimate $\nabla_{\alpha}g_{0}\nabla_{\alpha}g_{0}'$ and $\Sigma$.
    \item Do the singular value decomposition of both matrices and find the eigenvalues of every matrix. In general terms,
$\widehat{M}=\widehat{Q}'\widehat{\Lambda}\widehat{Q}$.
    \item Estimate the rank of $M$ using \begin{equation*}
        \widehat{r}_{M} = \#\{ \widehat{\lambda}_{i} \geq \frac{1}{n^b}, \hspace{0.2cm} i \in 1,...,m \},
    \end{equation*}
    where $\widehat{\lambda}_{i}$ are the elements of the diagonal matrix $\widehat{\Lambda}$ and $0<b<1$.
\end{enumerate}

\subsection{Estimation of Moore-Penrose g-inverse} \label{consistent_MOORE}
In this subsection, we discuss how to implement a consistent estimate of the optimal weighting matrix
\begin{equation}
    W=\left((I_{m}-\nabla_{\theta}g_{0})\widehat{\Sigma}
(I_{m}-\nabla_{\theta}g_{0})^{\prime}\right)^{\dagger}.
\end{equation}

The main problem in consistently estimating $W$ is that Moore-Penrose g-inverses are not continuous. An estimator based on the analog principle of $W$ might be inconsistent. Define 
$$A \equiv (I_{m}-\nabla_{\theta}g_{0})\Sigma
(I_{m}-\nabla_{\theta}g_{0})^{\prime},$$
$$\hat{A} \equiv (I_{m}-\hat{\nabla}_{\theta}g)\widehat{\Sigma}
(I_{m}-\hat{\nabla}_{\theta}g)^{\prime}.$$

\cite{stewart1969continuity} gives necessary and sufficient conditions to estimate consistently a Moore-Penrose g-inverse. Such conditions are: if $\hat{A} \xrightarrow{p} A$, then

\begin{equation*}
    \hat{A}^{\dagger} \xrightarrow{p} A^{\dagger}
    \iff Pr\left(Rank(\hat{A}) = Rank(A)\right)\xrightarrow{p}1.
\end{equation*}
To ensure that such a condition holds, we truncate the eigenvalues of $\widehat{A}$. First, use the spectral decomposition 
\begin{equation}
    \hat{A} = \hat{R}'\hat{P}\hat{R}.
\end{equation} 
Let $\hat{p}_{j}$ be the diagonal elements of $\hat{P}$, i.e, the eigenvalues. Then truncate the eigenvalues that are smaller than $\tfrac{1}{n^{b}}$, $\tilde{p}_{j} = \hat{p}_{j}\mathbbm{1}{ \{\hat{p}_{j}>\frac{1}{n^{b}}\}}$. 
Define $\Tilde{P}$ as the diagonal matrix composed by the $jth$ different elements $\tilde{p}_{j}$,  $\Tilde{P}^{\dagger}$ as the diagonal matrix with the $jth$ element given by $1/\tilde{p}_{j}$ when $\tilde{p}_{j}\neq0$, and zero otherwise, and compute
\begin{equation}
    \tilde{A} = \hat{R}\tilde{P}\hat{R}.
\end{equation}
Then, set 
 \begin{equation}
     \widehat{W} \equiv \tilde{A}^{\dagger} = \hat{R}\tilde{P}^{\dagger}\hat{R}.
 \end{equation}
Following the results of Theorem \ref{rank_estimation_theorem} $Pr\left(Rank(\tilde{A}) = Rank(A)\right)\xrightarrow{p}1$. Hence, the estimator $\widehat{W}$ satisfies \cite{stewart1969continuity} necessary and sufficient conditions for 
\begin{equation} \label{consistent moore}
    \widehat{W}\xrightarrow{p} W.
\end{equation}

\begin{enumerate}
    \item Compute the matrix $\hat{A} \equiv (I_{m}-\hat{\nabla}_{\theta}g)\widehat{\Sigma}
(I_{m}-\hat{\nabla}_{\theta}g)^{\prime}$.
 \item Get this spectral decomposition $\hat{A} = \hat{R}'\hat{P}\hat{R}.$
 \item Compute $\tilde{P}$.
 \item Compute the matrix $\tilde{A} =  \hat{R}\tilde{P}\hat{R} $.
 \item Set $\hat{W} = \tilde{A}^{\dagger} = \hat{R}\tilde{P}^{\dagger}\hat{R}$.
 
\end{enumerate}
\section{Monte Carlo} \label{MonteCarlo}
In the following subsections, we examine the finite sample properties of our method using Monte Carlo in two different applications. First, in a Static Bayesian Game proposed by \cite{pesendorfer2008asymptotic}. Second, in a setting that resembles the New-Keynesian model proposed by \cite{nakamura2018high}.

\subsection{Monte Carlo: Static Bayesian Game\label{MC}}
In this section, we demonstrate our methodology using an example from \cite{pesendorfer2008asymptotic}. We examine the properties of our test in finite samples, including size and power.
The example we consider is a static Bayesian Game involving two firms that must decide whether or not to enter a market simultaneously. The payoffs for Firm 1 and Firm 2 are as follows:
        $$\pi_{1}\left(a_{1}, a_{2}, s\right)= \begin{cases}\beta_{0}  \cdot s & \text { if } a_{1}=\text { active, } a_{2}=\text { not active } \\ \alpha_{1}\cdot s  & \text { if } a_{1}=\text { active, } a_{2}=\text { active } \\ 0 & \text { otherwise. }\end{cases}$$

        $$\pi_{2}\left(a_{1}, a_{2}, s\right)= \begin{cases}\alpha_{2}  \cdot s & \text { if } a_{2}=\text { active, } a_{1}=\text { not active } \\ \alpha_{3}\cdot s  & \text { if } a_{2}=\text { active, } a_{1}=\text { active } \\ 0 & \text { otherwise. }\end{cases}$$
The variable $s \in \{s_{1},s_{2},s_{3}\}$ is the state of the economy, with $s_{1}<s_{2}<s_{3}$. The utility for a firm $i$ with state $s$ and other firm's action $j$, is defined by:
\begin{equation*}
    u(a_{i},a_{j},s)=\pi_{i}(a_{i},a_{j},s)+\epsilon_{i},
\end{equation*}
where $\epsilon_{i}$ is a random unobserved variable distributed as a standard normal. Firm $i$th decides to be active whenever
\begin{equation*}
    u(a_{1},a_{2},s)=\pi_{i}(a_{1},a_{2},s)+\epsilon_{i}>0.
\end{equation*}
Firm $i$ has expectations about the probability of firm $j$ being active. In equilibrium, such expectations will be equal to the actual probability of firm $j$ choosing to be active, i.e
\begin{equation*}
    \theta_{1,s} = g\left((1-\theta_{2,s}) \cdot s\cdot  \beta_{0} + \theta_{2,s}\cdot\alpha_{1} \cdot s \right) \hspace{0.5cm} \forall s,
\end{equation*}
and
\begin{equation*}
    \theta_{2,s}=g\left((1-\theta_{1,s}) \cdot s\cdot  \alpha_{2} + \theta_{1,s})\cdot\alpha_{3} \cdot s \right) \hspace{0.5cm} \forall s,
\end{equation*}
where $\theta_{j,s}$ is the probability of choosing to be active for a firm $j$ given state $s$. The mapping $g$ is the CDF of a standard normal distribution. Our parameter of interest will be $\beta_{0}$, i.e, the coefficient of firm 1's payoff associated with being active when the other is not. The remaining parameters are considered nuisance parameters.

\subsubsection{Identification Problem}
In this example, we have six moment conditions (three states for every single firm) and four parameters, three of which are nuisance parameters. The Jacobian of the nuisance parameter is

\begin{equation}
    \nabla_{\alpha}g(\theta,\alpha,\beta)    
    \equiv \begin{pmatrix} \dot{g}_{1,1}\theta_{2,s_{1}}s_{1}& 0&0\\ \dot{g}_{1,2}\theta_{2,s_{2}}s_{2}& 0&0\\ \dot{g}_{1,3}\theta_{2,s_{3}}s_{3}& 0&0\\ 0 & \dot{g}_{2,1}(1-\theta_{1,s_{1}})s_{1} & \dot{g}_{2,1}\theta_{1,s_{2}}s_{1}  \\  
    0 & \dot{g}_{2,2}(1-\theta_{1,s_{2}})s_{2} & \dot{g}_{2,2}\theta_{1,s_{2}}s_{2} \\
    0 & \dot{g}_{2,3}(1-\theta_{1,s_{2}})s_{3} & \dot{g}_{2,3}\theta_{1,s_{3}}s_{3}, \\  
    \end{pmatrix} 
\end{equation}
Where the density of the standard normal, denoted by $\dot{g}_{i,s}$, is evaluated at $ (1-\theta_{i,s}) \cdot s\cdot \beta + \theta_{i,s}\cdot\alpha_{1} \cdot s$ for $s \in \{s_{1},s_{2},s_{3}\}$ and firm $i$. It is worth noting that when the choice probabilities of firm one are one-half for all possible states of the economy, i.e., $\theta_{1,s_{1}}=\theta_{1s_{2}}=\theta_{1,s_{3}}=\frac{1}{2}$, the Jacobian's rank becomes deficient.
\begin{equation}
    Rank(\nabla_{\alpha}g(\theta_{0},\alpha,\beta)) = 2 < q.
\end{equation}

This simple example illustrates how the degree of identification might depend on population values of the reduced form parameter, which are unknown.
\subsubsection{Monte Carlo exercise}
We test the hypothesis:
\begin{equation}
    H_{0}: \hspace{0.5cm}\beta = \beta_{0} \hspace{1.5cm} \textit{v.s} \hspace{1.5cm}H_{1}: \hspace{0.5cm}\beta \neq \beta_{0}.
\end{equation}

The value of $\beta_{0}$ is set to $\beta_{0} = 1.5$ in the identified case (case 1), and $\beta_{0} = 0.3$ in the not identified case (case 2). We compare three procedures: an Oracle test (Oracle) that assumes knowledge of the degrees of freedom $d$, our robust test (Robust) that estimates $d$, and the t-test (T-test) based on the asymptotic normality of the estimate of the structural parameter $\beta_{0}$ under point-identification. A summary of the performance of the different tests in terms of size for the identified case is presented in Table \ref{first bayesian game}. The findings indicate that rank estimation has a minimal impact on the procedure's overall level of uncertainty.

\begin{table}[h!]
\centering
\caption{\textbf{Size of the test at  5\%. Identified case. (Case 1): $Rank(\nabla_{\alpha}g)=3$}} 
\label{first bayesian game}
\begin{tabular}{lcccc}
\hline \hline
                     & $n=100$  & $n=250$  & $n=500$  & $n=1000$  \\  \hline
\textit{Oracle}             & 0.067 & 0.053 & 0.050 & 0.045   \\
\textit{Robust}               & 0.062 & 0.053 & 0.050 & 0.045    \\ 
\textit{T-test} & 0.055   & 0.047 & 0.046  & 0.049   \\
 \hline \hline
\end{tabular}
\end{table}
Table \ref{tab:tab2} considers the unidentified case, the T-test presents large size distortions with a larger size than $5\%$. In contrast, the Robust and Oracle tests control the size. Again, the estimation of the rank does not affect the performance of the Robust test.
 
\begin{table}[h!]
\centering
\caption{\textbf{Size of the test at  5\%. Not identified case. (Case 2): $Rank(\nabla_{\alpha}g)=2$} }
\label{tab:tab2}
\begin{tabular}{lcccc}
\hline \hline
                     & $n=100$  & $n=250$  & $n=500$  & $n=1000$  \\ \hline
\textit{Oracle}           & 0.058 & 0.051  & 0.052 & 0.050   \\ 
\textit{Robust} & 0.058   & 0.051 & 0.052  & 0.050   \\
\textit{T-test}               & 0.128 & 0.12  & 0.113 & 0.114    \\ 

\hline \hline
\end{tabular}
\end{table}
 
 In the following figures, we report the power of the different tests for the \textbf{$\textit{Case 2}$} of not identification under the null. In this example, the parameter of interest $\beta$ is identified under any possible alternative.
 \begin{figure} [h!]
 
  \centering
   \caption{ \textbf{Power of the test. (Case 2): Not identification under the null. }}

  \begin{subfigure}[b]{0.45\textwidth}
    \includegraphics[width=\textwidth]{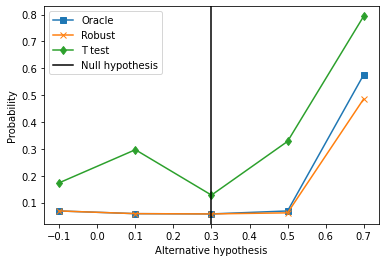}
    \caption{ \textbf{Sample size $n=100$} } 
    \label{fig:plot1}
  \end{subfigure}
  \hfill
  \begin{subfigure}[b]{0.45\textwidth}
    \includegraphics[width=\textwidth]{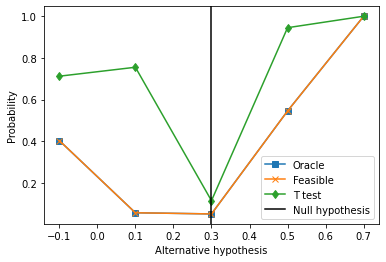}
    \caption{\textbf{Sample size $n=1000$}}
    \label{fig:plot2}
  \end{subfigure}

  \vspace{5pt} 

  \label{fig:two_plots}
\end{figure}

As a consequence of $\beta$ being identified under the alternative, the t-test has non-trivial power. Yet, the size is not controlled. The Robust and the Oracle tests yield similar results in terms of power, suggesting that the estimation of the degrees of freedom does not have a big impact in finite samples.

\subsection{Monte Carlo: Nakamura and Steinsson 2018 \label{MC nak}}
In this section, we carry out a Monte Carlo simulation based on \cite{nakamura2018high}. Our analysis consists of two parts. Firstly, we assess the finite sample properties by testing the validity of the calibrated (fixed) parameters in terms of both the size and power of our method. Secondly, we evaluate the finite sample properties of conducting inference on two structural parameters of interest, namely the Phillips Curve slope and the Information Effect. For these structural parameters, we compare the size and power obtained from our robust inference method with those from standard inference methods.

\cite{nakamura2018high} provides compelling evidence of the non-neutrality of monetary policy. They accomplish this by identifying the information effect of Federal Reserve announcements on economic fundamental beliefs. Specifically, \cite{nakamura2018high} utilizes data on unexpected changes in interest rates occurring within 30 minutes of a Fed announcement. In addition, they construct and estimate a structural model using the simulated method of moments. Their structural model involves the estimation of $m=33$ reduced-form parameters and $q=5$ structural parameters. The mapping function, $g(\cdot)$, has no closed-form solution, and it is therefore estimated numerically as in Section \ref{smm}.

Given that \cite{nakamura2018high} uses two different databases to estimate the reduced-form parameters, the matrix of variance and covariance $\Sigma$ is not identified. 

To illustrate our methodology, we employ a subset of moments and parameters, estimating $m=8$ reduced-form parameters and $q=2$ structural parameters. The latter corresponds to the slope of the Phillips Curve and the information effect of monetary policy. On the other hand, \cite{nakamura2018high} calibrate eight parameters in their study, while we calibrate $k=11$ parameters. Notably, the reduced-form parameters are estimated from a single database, enabling the identification of the variance and covariance matrix of the estimated parameters.

As demonstrated in Section \ref{AP}, the estimated rank of the Jacobian matrix is deficient, and the asymptotic variance matrix of the reduced-form parameters, $\widehat{\Sigma}$, is close to being singular. These limitations motivate the use of robust inference techniques. It is worth noting that throughout the Monte Carlo exercise, we employ the estimated $\hat{\Sigma}$ from Section \ref{AP} as the true one. Table \ref{tab:MD identified CI} presents the structural parameter values for different data-generating processes.

\begin{table}[h!] 
\centering
\begin{threeparttable}
\caption{\textbf{ Data Generating Process Structural Parameters.}}

\centering
\label{tab:MD identified CI}
\begin{tabular}{lcc}
    \hline    \hline          \textit{Notation}         &  \textit{Definition}  & \textit{Replication of Application } \\ \hline \hline

\multicolumn{3}{c}{\textit{Calibration (fixed parameters)} } \\ \hline

$\rho$     &  \textit{Subjective discount factor} &0.99      \\
$\eta$  &  \textit{Nominal rigidity} &  0.75 \\
$\omega$     &  \textit{Elasticity of marginal cost to output} &    2   \\
$\gamma$ & \textit{Coeff on lagged inflation} & 1   \\ 
$\phi$ & \textit{Endogenous feedback in Taylor rule} & 0.01 \\
$\pi$ & \textit{Inflation target shock} &  0 \\
$\delta$ & \textit{Elasticity of substitution across varieties} & 10  \\

$\sigma$ & \textit{Intertemporal el of substitution} &  0.50 \\ 

    $\rho_{1}$ & \textit{Autoregressive first root of
monetary shock} &  0.90 \\
$\rho_{2}$ & \textit{Autoregressive second root of 
monetary shock} &   0.79 \\
$b$ & \textit{Consumption Habit} &  0.95 \\
\hline

\multicolumn{3}{c}{\textit{Estimated parameters} } \\ \hline

$\kappa \zeta$ & \textit{Phillips Curve Slope} &  $1.0037\cdot10^{-4}$ \\
$\psi$ & \textit{Information Effect} &  $1.1617\cdot 10^{-6}$ \\

\\\hline  \hline
\end{tabular}
\begin{tablenotes}
\item \small \textbf{Note.} This DGP corresponds to the estimated values of the application in Section \ref{AP}.
\end{tablenotes}
\end{threeparttable}
\end{table}

\subsubsection{Calibration Validity}
We refer to calibration validity as not rejecting the calibrated value with our robust test. In other words, we test if $\beta=\beta_{0}$, where $\beta_{0}$ are the calibrated values. 

In this subsection, we test the validity of the calibration of the modified version of \cite{nakamura2018high} explained above. The vector $\beta$ is composed by all calibrated (fixed) parameters $$\beta = \left( \rho, \eta, \omega, \gamma, \phi, \pi, \delta, \sigma, \rho_{1}, \rho_{2}, b \right)'.$$

The nuisance parameters are $\alpha = \left(\kappa \zeta,  \psi \right)'$. There is no closed form solution for the mapping $g: \mathcal{B}\times\mathcal{A} \xrightarrow{} \Theta$, with $\mathcal{B} \subset \mathbb{R}^{11}$ and $\mathcal{A} \subset \mathbb{R}^{2}$. We use numerical methods to estimate $g(\cdot)$. We test the null hypothesis:
\begin{equation*}
    H_{0}: \beta = \beta_{0
    } \hspace{1cm} \textit{vs} \hspace{1cm} \beta \neq \beta_{0
    },
\end{equation*}
where $\beta_{0}$ are the fixed parameter values shown in Table \ref{tab:MD identified CI}.

\begin{table}[h!] 
\centering
\caption{\textbf{Size of the Robust test at  5\%.}} 
\label{size cal}
\begin{tabular}{lcccc}
\hline \hline
                     & $n=100$  & $n=250$  & $n=500$  & $n=1000$  \\   \hline 

\textit{Robust}                & 0.044 & 0.044 & 0.053 & 0.045   \\ 
\textit{Oracle}            & 0.007 & 0.007 & 0.009 & 0.003  \\     
 \hline \hline
\end{tabular}
\end{table}
In the replication of the application, the model is identified, albeit weakly. This is the reason why, in Table \ref{size cal}, it can be observed that using the oracle test (which utilizes the true degree of freedom) yields a size different from 5\%. The empirical size of the Robust test is satisfactory. To illustrate the power of the test, we maintain all elements of $\beta$ constant except for the shock to the inflation target, denoted as $\pi$. We focus on this alternative hypothesis since $\pi$ contributes the most to the deviation with maximum local power, as can be seen in Table \ref{contribution} in Section \ref{AP}.
\begin{figure} [h!]
 
  \centering
   \caption{ \textbf{Power of the robust test. Calibration Validity. }}

  \begin{subfigure}[b]{0.45\textwidth}
    \includegraphics[width=\textwidth]{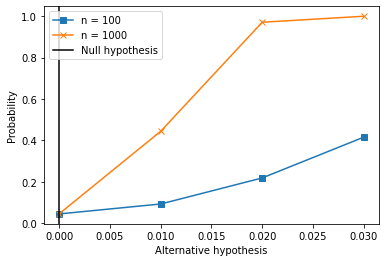}
    \label{fig:plot1}
  \end{subfigure}
  \vspace{5pt} 

  \label{fig:two_plots}
\end{figure}

\subsubsection{Inference on structural Parameters: The Phillips Curve Slope $\kappa \zeta$}

In this subsection, we assume our previous calibration is valid. Moreover, the vector of structural parameters of interest is $\beta = \kappa \zeta$, and the nuisance parameter is $\alpha = \psi$. We test the hypothesis:
\begin{equation}
    H_{0}: \kappa \zeta = \kappa\zeta_{0} \hspace{1cm} vs \hspace{1cm} H_{1}: \kappa \zeta \neq \kappa\zeta_{0},
\end{equation}
where $\kappa\zeta_{0}$ are the structural values presented in Table \ref{tab:MD identified CI} that correspond to every different case of study.

\begin{table}[h!]
\centering
\caption{\textbf{Size of the test at  5\%. Inference on Phillips Curve Slope $\kappa \zeta$ }} 
\label{tab:tab1}
\begin{tabular}{lcccc}
\hline \hline
              \textit{Test}       & $n=100$  & $n=250$  & $n=500$  & $n=1000$  \\ \hline 
                     \multicolumn{5}{c}{\textbf{Replication of Application}} \\ \hline
\textit{Oracle}                & 0.016 & 0.009 & 0.015 & 0.011   \\ 
\textit{Robust}             & 0.073 & 0.051 & 0.061 & 0.057   \\
\textit{T-test}               & 0 & 0 & 0 & 0    \\  \hline 

 \hline \hline
\end{tabular}
\end{table}

Under the Null hypothesis, both the t-test and the Oracle test do not exhibit a significance level of 5\%. This deviation from the expected significance level can be attributed to the weak identification of the model and the near singularity of $\Sigma$ (covariance matrix). In contrast, the robust test has a satisfactory size performance. Figure \ref{fig:two_plots} reports the results on power. The t-test has low power, while that of the Robust and Oracle tests is high for deviations from the right of the null hypothesis. In contrast to the Static Bayesian Game case, the model is unidentified under any alternative hypothesis. Furthermore, the covariance matrix $\Sigma$ is nearly singular. Consequently, the t-test fails to control the significance level. The Robust test has non-trivial power for any alternative hypothesis. Nevertheless, the power function is not symmetric around the null. Specifically, smaller alternative hypotheses relative to the null hypothesis yield lower power compared to larger alternatives. This would suggest that the derivative of the mapping $g$ with respect to $\kappa  \zeta$ is getting closer to 0 as $\kappa  \zeta$ goes to 0. 

\begin{figure} [h!]
 
  \centering
   \caption{ \textbf{Power of the test. Inference on Phillips Curve Slope estimate. }}

  \begin{subfigure}[b]{0.45\textwidth}
    \includegraphics[width=\textwidth]{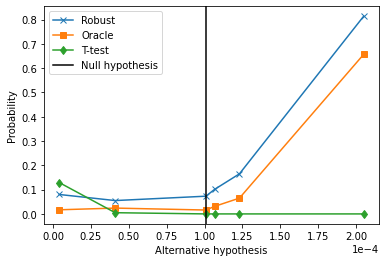}
    \caption{ \textbf{Sample size $n=100$} } 
    \label{fig:plot1}
  \end{subfigure}
  \hfill
  \begin{subfigure}[b]{0.45\textwidth}
    \includegraphics[width=\textwidth]{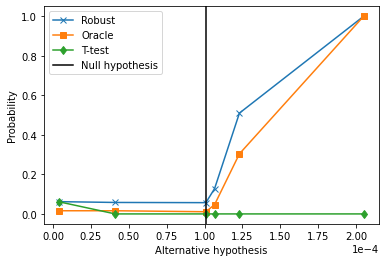}
    \caption{\textbf{Sample size $n=1000$}}
    \label{fig:plot2}
  \end{subfigure}

  \vspace{5pt} 

  \label{fig:two_plots}
\end{figure}

\subsubsection{Inference on Structural Parameters: The Information Effect $\psi$}

\begin{figure} [h!]
 
  \centering
   \caption{ \textbf{Power of the test. Inference on Information Effect $\psi$. }}

  \begin{subfigure}[b]{0.45\textwidth}
    \includegraphics[width=\textwidth]{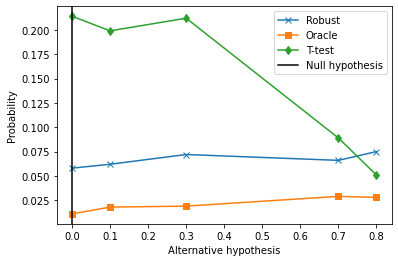}
    \caption{ \textbf{Sample size $n=100$} } 
    \label{fig:plot1}
  \end{subfigure}
  \hfill
  \begin{subfigure}[b]{0.45\textwidth}
    \includegraphics[width=\textwidth]{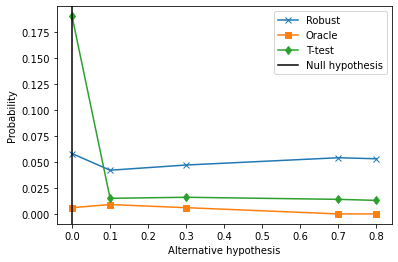}
    \caption{\textbf{Sample size $n=1000$}}
    \label{fig:plot2}
  \end{subfigure}

  \vspace{5pt} 

  \label{IE_power}
\end{figure}

In this subsection, the vector of structural parameters of interest is $\beta = \psi $, and the nuisance parameter is $\alpha = \kappa\zeta$. Whereas the rest of the structural parameters are assumed to be valid. We test the hypothesis:
\begin{equation}
    H_{0}: \psi= \psi_{0} \hspace{1cm} vs \hspace{1cm} H_{1}: \psi \neq \psi_{0},
\end{equation}
where $\psi_{0}$ are the structural values presented in Table \ref{tab:MD identified CI} that correspond to every different case of study. In the following Table \ref{IE size} and Figure \ref{IE_power} we report the results for this case. A remarkable difference relative to the previous section is that the Robust test has trivial power against any alternative, due to the fact that there exist observational equivalent models for any alternative $\beta_{1} \in \mathcal{B}$, i.e, $\forall$ $\beta_{1} \in \mathcal{B}$ $\beta_{1} \notin \mathcal{B}_{1}$, check Theorem \ref{feasible} for further details.

\begin{table}[h!]
\centering
\caption{\textbf{Size of the test at  5\%. Inference on Information Effect $\psi$ }} 
\label{IE size}
\begin{tabular}{lcccc}
\hline \hline
              \textit{Test}       & $n=100$  & $n=250$  & $n=500$  & $n=1000$  \\ \hline 
                     \multicolumn{5}{c}{\textbf{ Replication of Application}} \\ \hline
\textit{Oracle}                & 0.011 & 0.004 & 0.014 & 0.006   \\ 
\textit{Robust}             & 0.058 & 0.042 & 0.054 & 0.058   \\
\textit{T-test}               & 0.214 & 0.118 & 0.055 & 0.119    \\  

 \hline \hline
\end{tabular}
\end{table}

In this particular application, it seems that the Information Effect is not identified. This lack of identification could explain why there is trivial power in the Robust Test.

\section{Application: Inference in Nakamura and Steinsson 2018\label{AP}}In this section, we employ our robust inference method to re-analyze the study by \cite{nakamura2018high} and compare our results to those obtained through standard methods. In particular, first, we check the validity of the calibrated parameters in the main paper using our methodology. Second, we compare the inferences obtained by inverting our test with those obtained by standard inference methods. 

To evaluate the robustness of their approach, we estimate the rank of the Jacobian of the model evaluated at the estimated parameters, obtaining $\hat{r}_{\alpha}=Rank(\nabla_{\alpha}g(\widehat{\alpha}))=4<q=5$, indicating potential identification issues.

Unfortunately, in \cite{nakamura2018high} the asymptotic variance matrix of reduced form estimators is not identifiable due to the use of at least two different databases. To circumvent this issue, we opted to employ a subset of the estimated reduced form parameters, ensuring the identification of $\Sigma$. Our estimation procedure involves computing $m=8$ reduced-form parameters and $q=2$ structural parameters, the latter of which represent the slope of the Phillips Curve and the information effect of monetary policy. However, even though $\Sigma$ is identified in this simplified version of the problem, it is nearly singular. Table \ref{tab:shap} shows that the estimated rank of this matrix is $Rank(\hat{\Sigma})=5$. As a result, we advocate for identification-robust and singular-weighting-matrix-robust inference techniques in this application. The null hypothesis is
\begin{equation*}
    H_{0}: \beta = \beta_{0
    } \hspace{1cm} \textit{vs} \hspace{1cm}  H_{1}:\beta \neq \beta_{0
    },
\end{equation*}
where $\beta_{0} = \left( \rho, \eta, \omega, \gamma, \phi, \pi, \delta, \sigma, \rho_{1}, \rho_{2}, b \right)'$ are the values fixed in Table \ref{tab:MD identified CI}.

\begin{table}[h!] 
\caption{\textbf{Calibration validity}}

\centering
\label{tab:shap}
\begin{tabular}{lcccc}

\hline \hline                       & \textit{Statistic} & \textit{$Rank(\hat{\Sigma})$} & \textit{$Rank(\nabla_{\alpha}\hat{g})$} & \textit{Critical value} \\ \hline
\textit{Robust test}   & 3.18  & 5 &  1 & 9.487          \\ \textit{(Minimum Eigenvalue)} 
      &   & ($2.6114\cdot10^{-6}$) &  ($3.2299\cdot10^{-5}$) &           \\ \hline \hline    
\end{tabular}
\end{table} 
Table \ref{tab:shap} shows that the null hypothesis is not rejected by our robust test. Indicating that calibrated values are valid. Moreover, it is worth noting that the model is not identified, as the estimated value of $\widehat{r}_{\alpha} = Rank(\nabla_{\alpha}g(\widehat{\alpha}))$ is only one, which falls short of the required value of $q=2$. We conducted an empirical analysis of the local power and determined that the dimension of the local nontrivial power space, denoted by $\mathcal{B}_{\tau}$, is eight. This means that out of 11 parameters, only deviations in three parameters can lead to the rejection of the test with local nontrivial power.

\begin{table}[h!] 
\centering
\begin{threeparttable}
\caption{\textbf{ Contribution to Direction of Maximum Local Power.}}

\centering
\label{contribution}
\begin{tabular}{lcc}
    \hline    \hline          \textit{Notation}         &  \textit{Definition}  & \textit{Relative Weights}  \\ \hline \hline

\multicolumn{3}{c}{\textit{Calibration (fixed parameters)} } \\ \hline

$\rho$     &  \textit{Subjective discount factor} & 0.006      \\
$\eta$  &  \textit{Nominal rigidity} &  0.03 \\
$\omega$     &  \textit{Elasticity of marginal cost to output} &    0   \\
$\gamma$ & \textit{Coeff on lagged inflation} & 0.006 \\ 
$\phi$ & \textit{Endogenous feedback in Taylor rule} &  0.148\\
$\pi$ & \textit{Inflation target shock} &  0.674\\
$\delta$ & \textit{Elasticity of substitution across varieties} &  0 \\

$\sigma$ & \textit{Intertemporal el of substitution} &  0\\ 

    $\rho_{1}$ & \textit{Autoregressive first root of
monetary shock} & 0.13 \\
$\rho_{2}$ & \textit{Autoregressive second root of 
monetary shock} &  0.003  \\
$b$ & \textit{Consumption Habit} &   0.008

\\\hline  \hline
\end{tabular}
\end{threeparttable}
\end{table}
Table \ref{contribution} illustrates the individual contributions of each calibrated parameter towards the direction of maximum local power, as discussed in Remark \ref{relative weights}. The top three contributors to the maximum local power are the inflation target shock, the endogenous feedback in the Taylor rule, and the first autoregressive root of the monetary shock.

\begin{table}[h!] 
\caption{\textbf{Inference of structural parameters of interest}}

\centering
\label{tab:MD CI}
\begin{tabular}{llrlr}
\hline    \hline          $\beta$         &  \multicolumn{2}{c}{\textit{PC slope}}  & \multicolumn{2}{c}{\textit{Information effect}} \\ \hline
\textit{Estimator }  & \multicolumn{2}{c}{$1.0037\cdot10^{-4}$}   & \multicolumn{2}{c}{$1.1617\cdot10^{-6}$}  \\

\textit{CI T-test}      & (-9.246, & 9.295)   & (-0.5017 , & 0.5017)            \\
\textit{CI Robust test}      & ($2.1983\cdot10^{-5}$, & 0.0152)   & (0, & 0.9805)            \\
\textit{CI Bootstrap}      & ($4.087\cdot10^{-7}$, & 0.0321)   & ($1.7392\cdot10^{-4}$, & 0.9999)            \\\hline  \hline
\end{tabular}
\end{table}

Table \ref{tab:MD CI} presents a summary of the estimation and inference results for the structural parameters. In comparison, the confidence intervals (CI) based on the t-test are wider than those obtained using the bootstrap or the robust test. Moreover, the CI using the robust test appears to be shorter. CI bootstrap is the method that \cite{nakamura2018high} uses, and it is obtained using non-parametric bootstrap to estimate the finite sample distribution of the estimates.  Notably, the estimated information effect is approximately zero, indicating that nothing of the real effects of monetary policy stem from revisions in agents' fundamental economic beliefs. However, the estimation's confidence intervals are wide, primarily due to the exclusive use of inflation-related reduced-form parameters.

\section{Conclusion} \label{concl}

In this paper, we propose inference for structural parameters based on minimum distance that is robust to identification problems of nuisance parameters. Moreover, to implement this method is not necessary to have prior knowledge about the identification degree of the model. Last but not least, this inference method is computationally fast. Our robust inference methods can be applied to test the validity of calibrated (fixed) parameters in structural models under weak identification of nuisance parameters. They can be also applied to obtain identification-robust confidence intervals by inverting our robust test. Therefore, this method is suitable for a large class of applications in structural models commonly employed in Macroeconomics and Microeconomics.
\newpage
\appendix\renewcommand\thesection{Appendix \Alph{section}}\renewcommand
\thesubsection{\Alph{section}\arabic{subsection}}\renewcommand{\theequation
}{\Alph{section}-\arabic{equation}}\renewcommand{\thetheorem}{\Alph
{section}.\arabic{theorem}}\renewcommand{\theassumption}{\Alph{section}%
.\arabic{assumption}}\renewcommand{\theremark}{\Alph{section}\arabic
{subsection}.\arabic{remark}}%
%

\setcounter{assumption}{0}%

\section{Main Proofs\label{Main_Proofs}}

Define
\begin{equation}
f(\theta,\alpha):=\left(  \theta-g(\theta,\alpha,\beta_{0})\right)  ^{\prime
}W\left(  \theta-g(\theta,\alpha,\beta_{0})\right)  ,\label{f}%
\end{equation}%
\[
v(\theta):=\min_{\alpha\in\mathcal{A}}f(\theta,\alpha)
\]
and%
\[
\mathcal{M}(\theta)=\arg\min_{\alpha\in\mathcal{A}}f(\theta,\alpha).
\]
We shall use Theorem 4.2 in \cite{shapiro1985second}, which for the sake of reference is
repeated here adapted to our case.\bigskip

\noindent\textbf{Assumption S1}: There exists a constant $C\in\mathbb{R}$ such
that $v(\theta_{0})<C,$ and a compact set $\Pi$ such that $\{\alpha
\in\mathcal{A}:f(\theta,\alpha)\leq C\}\subset\Pi$ whenever $\theta$ is in a
neighborhood of $\theta_{0}.$\bigskip

\noindent\textbf{Assumption S2}: For every $\alpha_{0}\in\mathcal{M}%
(\theta_{0})\equiv\mathcal{A}_{0}$ there exists a local diffeomorphism (a
continuously differentiable mapping which is locally one-to-one, and with
inverse which is continuously differentiable) $\alpha=h(\gamma)$ such that the
function $t(\theta,\gamma):=f(\theta,h(\gamma))$ has Hessian matrix
non-singular at $(\theta_{0},\gamma_{0}),$ where $\gamma_{0}:=h^{-1}%
(\alpha_{0}),$ with respect to the variables $\gamma_{1},...,\gamma_{r},$ and
$t$ is independent of the variables $\gamma_{r+1},...,\gamma_{q}$ in a
neighborhood of $(\theta_{0},\gamma_{0}).$\bigskip

\begin{theorem} \label{4.2}  \textit{(\cite{shapiro1985second}, Theorem 4.2):}
Suppose $f(\theta,\alpha)$\textit{ is twice continuously differentiable,
}$\mathcal{M}(\theta_{0})$\textit{ is connected, and Assumptions S1 and S2
hold. Then, }$v(\theta)$\textit{ is twice continuously differentiable in a
neighborhood of }$\theta_{0}$\textit{ and }%
\begin{equation}
\nabla_{\theta}v(\theta_{0})=\nabla_{\theta}f(\theta_{0},\alpha_{0}%
)\label{4.6}%
\end{equation}

\begin{equation}
\nabla_{\theta\theta}^{2}v(\theta_{0})=\nabla_{\theta\theta}^{2}f(\theta
_{0},\alpha_{0})-\nabla_{\theta\alpha}^{2}f(\theta_{0},\alpha_{0})\left(
\nabla_{\alpha\alpha}^{2}f(\theta_{0},\alpha_{0})\right)  ^{\dagger}%
\nabla_{\theta\alpha}^{2}f(\theta_{0},\alpha_{0})^{\prime},\label{4.7}%
\end{equation}
\textit{where expressions in the right side of (\ref{4.6}) and (\ref{4.7})
give the same result for all }$\alpha_{0}\in M(\theta_{0}),$\textit{ and
(\ref{4.7}) is independent of a particular choice of the generalized
inverse}.
    
\end{theorem}
\noindent Since $f$ is continuous and $\mathcal{A}$ is compact, Assumption S1
holds. The following result follows from the well-known Rank Theorem (see e.g.
Fisher, 1966).

\begin{theorem} \label{rank}
\textit{All }$\alpha_{0}\in
\mathcal{A}_{0}$\textit{ are interior locally regular points of }$\mathcal{A}%
$\textit{. If }$r<q,$ \textit{and Assumption \ref{rank overparam} holds, i.e,} $Rank(\nabla_{\alpha}g_{0})= Rank(\nabla_{\alpha}g_{0}'W\nabla_{\alpha}g_{0})$ \textit{then Assumption S2 holds.}
\end{theorem}
\noindent\textbf{Proof of Theorem }\ref{distr}  By Assumption
\ref{differentiability}, $\mathcal{M}(\theta_{0})$ is non-empty, i.e.
$\alpha_{0}\in\mathcal{M}(\theta_{0})$. We allow for partial identification,
i.e. $\mathcal{M}(\theta_{0})$ containing more than one point. To prove the
theorem we shall apply Theorem \ref{4.2} above. From the previous discussion,
the conditions of the theorem are satisfied under our conditions. Simple calculus
shows that for the particular case in (\ref{f})
\begin{align*}
\nabla_{\theta}f(\theta_{0},\alpha_{0})  & =2\left(  \theta_{0}-g(\theta
_{0},\alpha_{0},\beta_{0})\right)  ^{\prime}W(I_{m}-\nabla_{\theta}g_{0})\\
& =0
\end{align*}
by the correct specification of the model, whereas
\begin{align*}
\nabla_{\theta\theta}^{2}f(\theta_{0},\alpha_{0})  & =2(I_{m}-\nabla_{\theta
}g_{0})^{\prime}W(I_{m}-\nabla_{\theta}g_{0})\\
\nabla_{\theta\alpha}^{2}f(\theta_{0},\alpha_{0})  & =-2(I_{m}-\nabla_{\theta
}g_{0})^{\prime}W\nabla_{\alpha}g_{0}\\
\nabla_{\alpha\alpha}^{2}f(\theta_{0},\alpha_{0})  & =2\nabla_{\alpha}g_{0}%
{}^{\prime}W\nabla_{\alpha}g_{0}.
\end{align*}
Write%
\begin{align*}
\widehat{F}(\beta_{0})  & =\min_{\alpha\in\mathcal{A}}n\left(  \widehat
{\theta}-g(\widehat{\theta},\alpha,\beta_{0})\right)  ^{\prime}\widehat
{W}\left(  \widehat{\theta}-g(\widehat{\theta},\alpha,\beta_{0})\right)  \\
& =n\left(  \widehat{\theta}-g(\widehat{\theta
},\widehat{\alpha},\beta_{0})\right)  ^{\prime}W\left(  \widehat{\theta}-g(\widehat
{\theta},\widehat{\alpha},\beta_{0})\right)  +\\
& n\left(  \widehat{\theta}-g(\widehat{\theta
},\widehat{\alpha},\beta_{0})\right)  ^{\prime}\left(  \widehat{W}-W\right)  \left(
\widehat{\theta}-g(\widehat{\theta},\widehat{\alpha},\beta_{0})\right)  .
\end{align*}
By the consistency of $\widehat{W},$ the second term on the right-hand size is
bounded by%
\begin{align*}
n\left\vert \left(  \widehat{\theta}-g(\widehat
{\theta},\widehat{\alpha},\beta_{0})\right)  \right\vert ^{2}\left\vert \widehat
{W}-W\right\vert  & =O_{P}(1)o_{P}(1)\\
& =o_{P}(1),
\end{align*}
where the $O_{P}(1)$ term follows from applying the Delta-Method and the
arguments above to the mapping $\theta\rightarrow\min_{\alpha\in\mathcal{A}%
}\left\vert \left(  \theta-g(\theta,\alpha,\beta_{0})\right)  \right\vert
^{2}.$ Then,%
\[
\widehat{F}(\beta_{0})=nv(\widehat{\theta})+o_{P}(1).
\]
By Theorem \ref{4.2} and the second order Delta-Method, we have%
\begin{align*}
\widehat{F}(\beta_{0})  & =n\left(  \widehat{\theta}-\theta_{0}\right)
^{\prime}\frac{1}{2}\nabla_{\theta\theta}^{2}v(\theta_{0})\left(
\widehat{\theta}-\theta_{0}\right)  +o_{P}(1).\\
& =Z^{\prime}UZ.
\end{align*}
Where
$
U:=(I_{m}-\nabla_{\theta}g_{0})^{\prime}\left(  W-W\nabla_{\alpha}g_{0}\left(
\nabla_{\alpha}g_{0}^{\prime}W\nabla_{\alpha}g_{0}\right)  ^{\dagger}%
\nabla_{\alpha}g_{0}^{\prime}W\right)  (I_{m}-\nabla_{\theta}g_{0}),
$ and $Z \sim N_{m}(0,\Sigma)$.

Now we find the distribution of $Z'UZ$. \cite{rao1972generalized} \textit{(P.171, Thm 9.2.1)} gives necessary and sufficient conditions for quadratic functions of correlated normal variables to be distributed as a chi-squared.
 In our context, $Z\sim N_{m}(0,\Sigma)$. If the condition $\Sigma U \Sigma U \Sigma = \Sigma U \Sigma $ holds, then $Z'UZ \sim \chi^{2}_{tr(U\Sigma)}$. We first define the following matrices:
 \begin{equation}
     E = \left(I_{m}-\nabla_{\theta
}g_{0} \right),
 \end{equation}
 \begin{equation}
     B =  \left(I_{m}-\nabla_{\alpha}g_{0}(\nabla_{\alpha}g_{0}'W\nabla_{\alpha}g_{0})\nabla_{\alpha}g_{0}' W \right),
 \end{equation}
 \begin{equation}
     C =  \left(I_{m}-W\nabla_{\alpha}g_{0}(\nabla_{\alpha}g_{0}'W\nabla_{\alpha}g_{0})\nabla_{\alpha}g_{0}' \right).
 \end{equation}
 Notice that,
\begin{equation} \label{equ}
    U=E'WBE=E'CWE.
\end{equation}
Using the two previous equalities in (\ref{equ}), and the fact that $W = \left(E\Sigma E'\right)^{\dagger}$ so that $ WE\Sigma E' W = W$, we can obtain
\begin{equation}
    \Sigma U \Sigma U \Sigma  = \Sigma E'CWE \Sigma E'WBE \Sigma = \Sigma E'CWBE \Sigma = \Sigma E'WBBE \Sigma =  \Sigma E'WBE \Sigma  = \Sigma U \Sigma.
\end{equation}
Using Assumptions \ref{rank overparam} and \ref{Non-deficiency} and the definition of $B$
\begin{equation}
    tr(U \Sigma) = tr(E'W E\Sigma)-tr(E'W \nabla_{\alpha}g_{0}\left(
\nabla_{\alpha}g_{0}^{\prime}W\nabla_{\alpha}g_{0}\right)  ^{\dagger}\nabla_{\alpha}g_{0}^{\prime}WE\Sigma)
\end{equation}
\begin{equation}
   = tr(W E\Sigma E')-tr(\nabla_{\alpha}g_{0}\left( 
\nabla_{\alpha}g_{0}^{\prime}W\nabla_{\alpha}g_{0}\right)  ^{\dagger}\nabla_{\alpha}g_{0}^{\prime}W)
\end{equation}
\begin{equation}
 =tr(W E\Sigma E')-tr(\left(
\nabla_{\alpha}g_{0}^{\prime}W\nabla_{\alpha}g_{0}\right)  ^{\dagger}\nabla_{\alpha}g_{0}^{\prime}W\nabla_{\alpha}g_{0})  
\end{equation}
\begin{equation}
 = Rank(E\Sigma E') - Rank(\nabla_{\alpha}g_{0}^{\prime}W\nabla_{\alpha}g_{0}) = r_{\Sigma} - r_{\alpha}.
\end{equation}

 \hfill\emph{Q.E.D.}\bigskip%

 \noindent\textbf{Proof of Theorem }\ref{feasible equal oracle}:
 Let the feasible test be 
\begin{equation}
    \tilde{\phi}_{\tau} = 1(\widehat{F}(\beta_{0}%
)>\chi_{d}^{2})
\end{equation}
and the oracle test
\begin{equation}
     \hat{\phi}_{\tau} = 1(\widehat{F}(\beta_{0}%
)>\chi_{\widehat{d}}^{2}).
\end{equation}
The power function is defined as 
\begin{equation}
    \pi_{\tau}(\beta)\equiv \lim_{n\rightarrow\infty
}\mathbb{P}_{\beta}\left(  \tilde{\phi}_{\tau}=1\right)
\end{equation}

\begin{equation*}
    \mathbb{P}_{\beta}\left(  \hat{\phi}_{\tau}=1\right) = \sum_{j=1}^{m}\mathbb{P}_{\beta}\left(  \widehat{F}(\beta_{0}%
)>\chi_{j}^{2}\right) \mathbb{P}_{\beta}(\widehat{d} = j) 
\end{equation*}
\begin{equation} \label{prob}
    = \mathbb{P}_{\beta}\left(  \widehat{F}(\beta_{0}%
)>\chi_{d}^{2}\right) \mathbb{P}_{\beta}(\widehat{d} = d)+ \sum_{j=1, \hspace{0.1cm} j\neq d}^{m}\mathbb{P}_{\beta}\left(  \widehat{F}(\beta_{0}%
)>\chi_{j}^{2}\right) \mathbb{P}_{\beta}(\widehat{d} = j),
\end{equation}

by consistency of the estimated degree of freedom $\widehat{d}$, $lim_{n\rightarrow\infty
}\mathbb{P}_{\beta}(\widehat{d} = d)= 1$, equation (\ref{prob}) equals

\begin{equation}
    = \mathbb{P}_{\beta}\left(  \widehat{F}(\beta_{0}%
)>\chi_{d}^{2}\right) \mathbb{P}_{\beta}(\widehat{d} = d) + o_{P}(1) = \mathbb{P}_{\beta}\left(  \tilde{\phi}_{\tau}=1\right) + o_{P}(1).
\end{equation}

 \hfill\emph{Q.E.D.}\bigskip%

 \noindent\textbf{Proof of Theorem }\ref{power}: Let $\alpha_{1}$ and $\alpha_{0}$ be the plim of

\begin{equation} \label{altern}
    \widehat{\alpha}_{1} = argmin_{\alpha \in \mathcal{A}}\left(\widehat{\theta}-g(\widehat{\theta},\alpha,\beta_{1})\right)'\widehat{W}\left(\widehat{\theta}-(g(\widehat{\theta},\alpha,\beta_{1})\right) +  \lambda
_{n}\left\vert \alpha\right\vert ^{2},
\end{equation}
and \begin{equation} \label{altern}
    \widehat{\alpha}_{0} = argmin_{\alpha \in \mathcal{A}}\left(\widehat{\theta}-g(\widehat{\theta},\alpha,\beta_{0})\right)'\widehat{W}\left(\widehat{\theta}-(g(\widehat{\theta},\alpha,\beta_{0})\right) +\lambda
_{n}\left\vert \alpha\right\vert ^{2},
\end{equation}
respectively. Define $g(\widehat{\theta
},\widehat{\alpha}_{1},\beta_{1}) \equiv \hat{g}_{1}$, and $g(\widehat{\theta
},\widehat{\alpha}_{0},\beta_{0}) \equiv \hat{g}_{0}$. From now on we assume we are under the alternative hypothesis $\beta_{1}$. Using the definition of $\hat{F}(\beta_{0})$,
\begin{align*}
\widehat{F}(\beta_{0})  & =\min_{\alpha\in\mathcal{A}}n\left(  \widehat
{\theta}-g(\widehat{\theta},\alpha,\beta_{0})\right)  ^{\prime}\widehat
{W}\left(  \widehat{\theta}-g(\widehat{\theta},\alpha,\beta_{0})\right)  \\
& = n\left(  \widehat
{\theta}-\hat{g}_{0}\right)  ^{\prime}
W\left(  \widehat{\theta}-\hat{g}_{0}\right)\left(1+o_{p}(1)\right)\\
& = n\left(  \widehat
{\theta}-\hat{g}_{1}+\hat{g}_{1}-\hat{g}_{0}\right)  ^{\prime}
W\left(  \widehat{\theta}-\hat{g}_{1}+\hat{g}_{1}-\hat{g}_{0}\right)\left(1+o_{p}(1)\right)\\
& = n\left(  \widehat
{\theta}-\hat{g}_{1}+\hat{g}_{1}-\hat{g}_{0}\right)  ^{\prime}
W\left(  \widehat{\theta}-\hat{g}_{1}+\hat{g}_{1}-\hat{g}_{0}\right)\left(1+o_{p}(1)\right) \\
& = [n\left(  \widehat
{\theta}-\hat{g}_{1}\right)  ^{\prime}
W\left(  \widehat{\theta}-\hat{g}_{1}\right)+2n\left(  \hat{g}_{1}-\hat{g}_{0}\right)  ^{\prime}
W\left(  \widehat{\theta}-\hat{g}_{1}\right)+\left(  \hat{g}_{1}-\hat{g}_{0}\right)  ^{\prime}
W\left(  \hat{g}_{1}-\hat{g}_{0}\right)]\left(1+o_{p}(1)\right)\\
& = [O_{P}(1)+2O_{P}(\sqrt{n})+O_{P}(n)]\left(1+o_{P}(1)\right)\\
&  = O_{P}(n).
\end{align*}
Last equality is true whenever $ |g(\theta,\alpha_{1},\beta_{1}) - g(\theta,\alpha_{0},\beta_{0})|>0$. Therefore,
\begin{equation}
    min_{\alpha \in \mathcal{A}}|g(\theta_{0}
,\alpha,\beta_{1})-g(\theta_{0},\alpha_{0},\beta_{0})|>0
\end{equation}
is a sufficient condition for the third term to be $O_{P}(n)$, and hence, the test to have power against $\beta = \beta_{1}$.

 \hfill\emph{Q.E.D.}\bigskip%

 \noindent\textbf{Proof of Theorem } \ref{power local}.
From proof of Theorem \ref{power} we have that under Pittman alternatives, and noticing that $\sqrt{n}\left(  \hat{g}_{1}-\hat{g}_{0}\right) = \widehat{\nabla}_{\beta} g_{0}\sqrt{n}(\beta_{1}-\beta_{0})=\widehat{\nabla}_{\beta} g_{0}\delta = \nabla_{\beta}g_{0}\delta + o_{P}(1)$. 

\begin{equation} 
\widehat{F}(\beta_{0}) = n\left(  \widehat
{\theta}-\hat{g}_{1}\right)  ^{\prime}
W\left(  \widehat{\theta}-\hat{g}_{1}\right)+2\delta'\nabla_{\beta}g_{0} ^{\prime}  
W \sqrt{n} \left(  \widehat{\theta}-\hat{g}_{1}\right)+  \delta'\nabla_{\beta}g_{0} ^{\prime}
W\nabla_{\beta}g_{0}\delta + o_{P}(1).
\end{equation}

After noticing that the limit of this previous expression under the alternative can be represented as:
\begin{equation}
    \lim_{n\xrightarrow{}\infty} \widehat{F}(\beta_{0}) = Y'WY+b'W Y+c,
\end{equation}

where $ \sqrt{n}\left(  \widehat
{\theta}-\hat{g}_{1}\right) \xrightarrow{d} Y \sim N_{m}(0, E\Sigma E')$, $b = W\nabla_{\beta}g_{0}\delta $ and $c = \delta'\nabla_{\beta}g_{0} ^{\prime}
W\nabla_{\beta}g_{0}\delta$.
This quadratic form holds the necessary and sufficient conditions of \cite{rao1972generalized} to be a non-central chi-squared distribution, $\chi^{2}(d,k)$, where $d$ are the degrees of freedom and $k$ is the non-centrality parameter. The value of $k$ can be found by

\begin{equation}
    k \equiv b'E \Sigma E' W E \Sigma E' b  
\end{equation}
After substituting the value of $b$ into previous equation, and rearranging terms,
\begin{equation}
    k = \delta' \nabla_{\beta}g_{0}'W\nabla_{\beta}g_{0}\delta.
\end{equation}

A non-central chi-squared distribution $\chi^{2}(d,k')$ stochastically dominates $\chi^{2}(d,k)$, with $k' > k$. Hence, If we are under the Pittman alternative, and the direction $\delta$ is such that $k > 0$, the power function will satisfy $\pi_{\tau}(\beta_{1}) > \pi_{\tau}(\beta_{0}) = \tau$, after noticing that the probability distribution of $\pi_{\tau}(\beta_{0})$ follows a chi-squared distribution with $k=0$.

 \hfill\emph{Q.E.D.}\bigskip%

 \noindent\textbf{Proof of Theorem }\ref{rank_estimation_theorem}:
 Recall that $\widehat{M}$ and $M$ are of dimension $m\times m$. Let $r \equiv Rank(M)$. Assume eigenvalues of $\widehat{M}$ are in descending order. Define the following sequence of events:
  \begin{equation}
      R_{n} \equiv \{\widehat{r}.
      =r\},
  \end{equation}
 Let $\hat{\lambda}_{i}$ and $\lambda_{i}$ be the $ith$ eigenvalue of $\widehat{M}$ and $M$, respectively. Let

    \begin{equation}
      E_{n} \equiv \{\widehat{\lambda}_{r
      }>\frac{1}{n^b}
      \},
  \end{equation}

    \begin{equation}
      F_{n} \equiv \{\widehat{\lambda}_{r+1
      }< \frac{1}{n^b}
      \},
  \end{equation}
  The idea of the proof is to show that $Pr(E_{n})\xrightarrow{} 1$, $Pr(F_{n})\xrightarrow{} 1$ as $n$ grows to infinite, and hence, 
  \begin{equation}
      Pr(E_{n} \cap F_{n}) = Pr(E_{n}) + Pr(F_{n}) - Pr(E_{n} \cup F_{n}) \xrightarrow{} 1.
  \end{equation}
Since $E_{n
 }\cap F_{n} \subset R_{n}$, $Pr(R_{n})\geq Pr(E_{n} \cap F_{n}) \xrightarrow{}1$.\\
Recall assumption \ref{norm_rank},
    \begin{equation}\label{bounded}
       \sqrt{n} \left(Vec(\widehat{A})-Vec(A)\right)\xrightarrow{d} N_{m^{2}}(0, \Omega).
    \end{equation}
    Define
    \begin{equation}\label{vectorization}
        \widehat{B} \equiv \widehat{A} -A,
    \end{equation}
    the element of row $ith$ and column $jth$ of $\widehat{B}$ is defined by $b_{ij}$. By (\ref{bounded}) all the elements $b_{i,j}$ are $O_{P}(\frac{1}{\sqrt{n}})$. Define $\hat{\mu}_{l}$ as the $lth$ eigenvalue of the matrix $\widehat{B}$. By definition $ ||\widehat{A}-A ||_{2}=\max_{l\in \{1,..,m\}}\hat{\mu}_{l}$. Using the Gershgorin circle theorem, the maximum eigenvalue of the matrix will be bounded by the disc with the largest radius, i.e.,
    
    \begin{equation} \label{Bounded}
    ||\widehat{A}-A ||_{2}\equiv\max_{l\in \{1,..,m\}}\hat{\mu}_{l} \leq \max_{i \in \{1,..,m\}}\sum_{j\neq i}^{q} |b_{ij}|= O_{P}\left(\frac{1}{\sqrt{n}}\right).
    \end{equation} 
    The last equality comes from the fact that all the elements of (\ref{vectorization}) are bounded in probability at a rate $\sqrt{n}$. Equation (\ref{Bounded}) implies
    \begin{equation}
    ||\widehat{A}-A ||_{2} = O_{P}\left(\frac{1}{\sqrt{n}}\right).
    \end{equation}
Now we are going to use the fact that estimated eigenvalues whose population eigenvalue is 0, converge to $0$ at a rate of $n^{-1}$. Under Assumption \ref{norm_rank},
\begin{equation} \label{supercons}
    \widehat{\lambda}_{l} = O_{P}\left(\frac{1}{n}\right) \hspace{0.5cm} \textit{for }r<l\leq m,
\end{equation}
as follows from Theorem \ref{RS proof}.

Suppose the case of $r=0$, notice $\lambda_{l}=0$ for $l\in\{1,..,m\}$, and  $A=0$. Then by result (\ref{supercons})
    \begin{equation}
        Pr(R_{n})=P(\max_{l\in\{1,...,m\}} \hat{\lambda}_{l} \leq \frac{1}{n^b})  \xrightarrow{n \xrightarrow{} \infty} 1,
    \end{equation}
    Suppose the case of $r>0$. Let $\Bar{c}=\frac{\lambda_{r}}{2}>0$. By Weyl's perturbation theorem and (\ref{Bounded}):
    \begin{equation}
        \max_{l\in\{1,...,q\}} |\hat{\lambda}_{l}-\lambda_{l}| \leq  ||\widehat{A}-A ||_{2} = o_{P}(1).
    \end{equation}
notice that for $n$ large enough we have $\frac{1}{n^b}<\bar{c}$ and using (\ref{Bounded}),
    \begin{equation*}
        Pr(E_{n}) = Pr(\hat{\lambda}_{r} \geq \frac{1}{n^b}) \geq Pr(\hat{\lambda}_{r} > \Bar{c}) = Pr(\hat{\lambda}_{r}-2\bar{c} > -\bar{c}) = Pr((\hat{\lambda}_{r}-\lambda_{r}) > -\bar{c}) =
    \end{equation*}
    \begin{equation}\label{consistency rank}
        = Pr(-(\hat{\lambda}_{r}-\lambda_{r}) < \bar{c})\geq Pr(|(\hat{\lambda}_{r}-\lambda_{r})| < \bar{c})\geq Pr(\max_{l\in \{1,..,m\}}|(\hat{\lambda}_{l}-\lambda_{l})| < \bar{c})\xrightarrow{n\xrightarrow{}\infty} 1.
    \end{equation}
    Suppose $r=m$. Then, by (\ref{consistency rank}) and noticing that $E_{n} \subset R_{n}$, $Pr(R_{n})\xrightarrow{n\xrightarrow{}\infty}1$.
    Now, assume $r<m$, Recall $F_{n}\{\hat{\lambda}_{r+1}\leq \frac{1}{n^b}\}$. By (\ref{supercons}) and noticing that $\lambda_{l}=0$ $\forall l>r$,  we have:
    \begin{equation}
        Pr(F_{n})=Pr(\hat{\lambda}_{r+1}\leq \frac{1}{n^b})\xrightarrow{n\xrightarrow{}\infty}1.
    \end{equation}
    Therefore $ Pr(E_{n})\xrightarrow{n\xrightarrow{}\infty}1$ and $ Pr(F_{n})\xrightarrow{n\xrightarrow{}\infty}1$ which implies $ Pr(R_{n})\xrightarrow{n\xrightarrow{}\infty}1$.
    \hfill\emph{Q.E.D.}\bigskip%

 \noindent\textbf{Proof of Theorem }\ref{RS proof}  \textit{(\cite{robin2000tests}, Theorem 3.1 and Theorem 3.2)}.
 Let \(K\) be a Gram matrix \(K = MM'\), where \(M\) is an \(m \times q\) matrix. Perform the singular value decomposition of \(M\), \(M = C\Lambda D'\), where every column of \(C\) and \(D\) are eigenvectors of \(MM'\) and \(M'M\) respectively. \(\Lambda\) is a diagonal matrix of eigenvalues. We order the eigenvalues in decreasing order. We partition \(C = [C_{r_M}, C_{m-r_M}]\) and \(D = [D_{r_M}, D_{q-r_M}]\) where \(C_{m-r_M}\) and \(D_{q-r_M}\) are the eigenvectors associated with eigenvalues \(\lambda_i = 0\). The matrices \(C\) and \(D\) have the following properties: \(C'C = I_m\), \(D'D = I_q\), \(C_{r_M}'C_{m-r_M} = 0\), \(D_{r_M}'D_{q-r_M} = 0\).\\
Define the following $m-r_{M} \times m-r_{M}$ matrix $\hat{A} \equiv C_{m-r_{M}}'(\hat{M}-M)D_{q-r_{M}}$, 
The proof is divided into three steps. First, we prove that the eigenvalues of the matrix $\hat{A}\hat{A}'$, denoted as $\tilde{\lambda}_{i}$ are $O_{p}(n^{-1})$. Second, we prove that  $o_{p}(1)=|n\hat{A}\hat{A}'-n\hat{\lambda}_{i}I_{m-r_{M}}|$ where $|\cdot|$ is notation for the determinant, and $\hat{\lambda}_{i}$ are the eigenvalues of $\hat{K} = \hat{M}\hat{M}'$ for $i=r_{M}+1,...,m$. In other words, we prove that the determinantal function when evaluated at $\hat{\lambda}_{i}$ is $o_{p}(1)$. Third, we prove that $o_{p}(1)=|n\hat{A}\hat{A}'-n\hat{\lambda}_{i}I_{m-r_{M}}|$ implies that $n\hat{\lambda}_{i} = n\tilde{\lambda}_{j}+o_{p}(1)$, for some $i=r_{M}+1,...,m$ and $j =1,..,m-r_{M}$, and hence, $\hat{\lambda}_{i} = O_{p}(n^{-1})$.\\

First part of the proof. By Assumption \ref{norm_rank} and using the Delta method,
\begin{equation}
    \sqrt{n}Vec(\hat{A}) \xrightarrow{d}N(0,(C_{m-r_{M}}'\otimes D_{q-r_{M}}')\Omega (C_{m-r_{M}}\otimes D_{q-r_{M}})),
\end{equation}
therefore, by the continuous mapping theorem, $nVec(\hat{A})'Vec(\hat{A}) = O_{p}(1)$. Moreover, 
\begin{equation} \label{trace}
    nVec(\hat{A})'Vec(\hat{A}) = nTrace(\hat{A}\hat{A}') = n\sum_{i = 1}^{m-r_{M}} \tilde{\lambda}_{i} = O_{p}(1).
\end{equation}
Since $\tilde{\lambda}_{i}$ is non-negative, equation (\ref{trace}) implies $n\tilde{\lambda}_{i} = O_{p}(1)$.

Second part of the proof. Represent $\hat{M}$ as $\hat{M} = M+(\hat{M}-M) $ notice that, 
\begin{equation} \label{first}
    C_{r_{M}}'\hat{M} = C_{r_{M}}'(C\Lambda D'+\hat{M}-M) = \Lambda_{r_{M}}D_{r_{M}}'+C_{r_{M}}'(\hat{M}-M) = \Lambda_{r_{M}}D_{r_{M}}'+O_{p}(n^{ -\sfrac{1}{2}}),
\end{equation}
where $\Lambda_{r_{M}}$ is the eigenvalue matrix of $r$ non-zero eigenvalues of $M$. The second equality  in (\ref{first}) comes from the orthogonality of $C_{r_{M}}$ with $C_{m-r_{M}}$ and $C_{r_{M}}'C_{r_{M}} = I_{r_{M}}$. On the other hand,
\begin{equation} \label{second}
    C_{m-r_{M}}'\sqrt{n}\hat{M}= C_{m-r_{M}}'\sqrt{n}(\hat{M}-M),
\end{equation}
after noticing that $C_{m-r_{M}}'M = 0$ by $C_{m-r_{M}}'C_{r_{M}}'=0$, and that the last $m-r_{M}$ elements of $\Lambda$ being 0. By $\hat{\lambda}_{i}$ being an eigenvalue of $\hat{M}\hat{M}'$ we have that:
\begin{equation*}
    \begin{split}
    0 &= |\hat{M}\hat{M}'-\hat{\lambda}_{i}I_{m}| = |(C_{r_{M}}', C_{m-r_{M}}'\sqrt{n})(\hat{M}\hat{M}'-\hat{\lambda}_{i}I_{m}) (C_{r_{M}}, C_{m-r_{M}}\sqrt{n})| \\
    &=  |\begin{pmatrix}
    C_{r_{M}}'\hat{M}\hat{M}'C_{r_{M}}  & C_{r_{M}}'\sqrt{n}\hat{M}\hat{M}'C_{m-r_{M}} \\
    C_{m-r_{M}}'\sqrt{n}\hat{M}\hat{M}'C_{r_{M}} &  C_{m-r_{M}}'n\hat{M}\hat{M}'C_{m-r_{M}}
\end{pmatrix}|  \\
    &= |\begin{pmatrix}
\Lambda_{r_{m}}\Lambda_{r_{m}}'+o_{p}(1)  & \Lambda_{r_{m}}D_{r_{M}}'\sqrt{n}(\hat{M}-M)'C_{m-r_{M}} + o_{p}(1)\\
    C_{m-r_{M}}'\sqrt{n}(\hat{M}-M)D_{r_{M}}\Lambda_{r_{m}}'  o_{p}(1)&  C_{m-r_{M}}'n(\hat{M}-M)(\hat{M}-M)'C_{m-r_{M}} 
\end{pmatrix} - \hat{\lambda}_{i}\begin{pmatrix}
    I_{r_{M}}  & 0 \\
    0 & nI_{m-r_{M}}
\end{pmatrix}| 
\end{split}
\end{equation*}
where the last equality comes from (\ref{first}) and (\ref{second}). Using the continuity of a determinant 
\begin{equation*}
\begin{split}
&= |\begin{pmatrix}
\Lambda_{r_{m}}\Lambda_{r_{m}}'  & \Lambda_{r_{m}}D_{r_{M}}'\sqrt{n}(\hat{M}-M)'C_{m-r_{M}} \\
    C_{m-r_{M}}'\sqrt{n}(\hat{M}-M)D_{r_{M}}\Lambda_{r_{m}}'  &  C_{m-r_{M}}'n(\hat{M}-M)(\hat{M}-M)'C_{m-r_{M}} 
\end{pmatrix}
- n\hat{\lambda}_{i}
\begin{pmatrix}
    0  & 0 \\
    0 & I_{m-r_{M}}
\end{pmatrix}| + o_{p}(1).
\end{split}
\end{equation*}
Now we use the fact that $|\begin{pmatrix}
    F & G \\ H & J
\end{pmatrix}| = |F||J-HF^{-1}G|$, for $F$ nonsingular, and $D_{q-r_{M}} = I_{m-r_{M}}-D_{r_{M}}$
\begin{equation*}
    \begin{split}
        0&= |\Lambda_{r_{m}}\Lambda_{r_{m}}'||C_{m-r_{M}}'n(\hat{M}-M)D_{q-r_{M}}D_{q-r_{M}}'(\hat{M}-M)'C_{m-r_{M}} -n \hat{\lambda}_{i}I_{m-r_{M}}| +o_{p}(1).
    \end{split}
\end{equation*}
Therefore, 
\begin{equation}
    |C_{m-r_{M}}'n(\hat{M}-M)D_{q-r_{M}}D_{q-r_{M}}'(\hat{M}-M)'C_{m-r_{M}} -n \hat{\lambda}_{i}| = o_{p}(1).
\end{equation}
Third part of the proof. We use the representation of the determinant as the roots of the characteristic polynomial evaluated at $\hat{\lambda}_{i}$:
\begin{equation} \label{characteristic pol}
    |C_{m-r_{M}}'n(\hat{M}-M)(\hat{M}-M)'C_{m-r_{M}} -n \hat{\lambda}_{i}| = (n\tilde{\lambda}_{1}-n\hat{\lambda}_{i})(n\tilde{\lambda}_{2}-n\hat{\lambda}_{i})\cdot...\cdot (n\tilde{\lambda}_{p-r_{M}}-n\hat{\lambda}_{i}) = o_{p}(1).
\end{equation}
Equation $(\ref{characteristic pol})$ implies that $(n\tilde{\lambda}_{j}-n\hat{\lambda}_{i}) = o_{p}(1)$ for some $j = 1,...,m-r_{M}$. Hence, using the result obtained in the first step, $n\tilde{\lambda}_{j} = O_{p}(1)$, and hence
\begin{equation}
    n\hat{\lambda}_{i} = O_{p}(1).
\end{equation}

 \hfill\emph{Q.E.D.}\bigskip%



\newpage
\phantomsection
\bibliographystyle{ectabib}
\bibliography{bibliography}

\end{document}